\documentclass[prb,preprint,amsmath,amssymb,amsfonts,citeautoscript,aps,superscriptaddress]{revtex4-2}  
\usepackage{times,graphicx,epstopdf,dcolumn,bm,color,braket,multirow,xfrac}
\usepackage[hidelinks,colorlinks=true,allcolors=blue]{hyperref}
\setcitestyle{super}

\begin{document} 

\author{Sung-Hoon Lee}  \email{lsh@khu.ac.kr}
\affiliation{Department of Applied Physics, Kyung Hee University, Yongin, Korea}

\author{Ki-Ha Hong}  \email{kiha.hong@hanbat.ac.kr}
\affiliation{Department of Materials Science and Engineering, Hanbat National University, Daejeon, Korea}

\title{Crystal structure and collective oxygen transport in high-temperature Ta$_\mathbf{2}$O$_\mathbf{5}$}

\begin{abstract} 
	
Ionic conduction in crystalline solids is conventionally understood to proceed via atomic-scale defects such as vacancies or interstitials. Here, by addressing the long-standing structural ambiguity of high-temperature tetragonal tantalum pentoxide (H-Ta$_2$O$_5$), we identify a qualitatively different transport mechanism. Based on first-principles calculations, we propose that H-Ta$_2$O$_5$ adopts a chiral framework composed of orthorhombic building units interconnected by screw-rotation planes, with a tantalum sublattice consistent with available transmission electron microscopy observations. Our ab initio molecular dynamics simulations reveal collective, one-dimensional oxygen migration within this stoichiometric lattice at temperatures of a few hundred degrees Celsius. This cooperative transport is enabled by the structural flexibility of octahedral coordination at the screw-rotation planes, which allows extensive lattice relaxation and dynamic charge redistribution, yielding a migration barrier of $\sim$0.2 eV. These results provide a microscopic interpretation of the reported high and anisotropic oxygen conductivity in H-Ta$_2$O$_5$.

\end{abstract}

\maketitle

\newpage

\subsection*{\normalsize Introduction}   

Crystalline solids are characterized by highly ordered atomic arrangements within repeating unit cells, forming well-defined crystal lattices. Conventionally, atoms are assumed to remain near their equilibrium positions below melting or sublimation temperatures, undergoing primarily localized vibrations. Within this framework, ionic conduction in crystalline materials is typically understood to proceed via defects such as vacancies, interstitials, or partial occupancies \cite{farrington1979, knauth2002, sau2024, jun2024, sotoudeh2024, jacobson2010, tuller2017, dwivedi2020}. 
For example, solid fast-ion conductors used in lithium-battery applications often rely on partially occupied Li sites that facilitate Li-ion migration \cite{sau2024, jun2024, sotoudeh2024}. In solid oxide fuel cells, doping introduces oxygen vacancies that act as the primary carriers for oxide-ion transport \cite{jacobson2010, tuller2017, dwivedi2020}. 
Here, we provide theoretical evidence that challenges this conventional picture by showing that the high-temperature tetragonal phase of Ta$_2$O$_5$ can support collective oxygen transport channels within a stoichiometric lattice.

Ta$_2$O$_5$, owing to its high dielectric constant and wide bandgap, is technologically important for applications including high-$k$ dielectrics \cite{brennecka2006, robertson2015, wang2018}, resistive random-access memory \cite{lee2011, kim2020, wang2023}, photo- and electrocatalysis \cite{gurylev2022, zheng2022}, and antireflection coatings \cite{sertel2019}. It primarily exists in two crystalline phases: the orthorhombic L-Ta$_2$O$_5$ at lower temperatures and the tetragonal H-Ta$_2$O$_5$ at higher temperatures. The transition between these phases occurs via an enantiotropic and sluggishly reversible transformation at $\sim$1360~°C \cite{zaslavskii1955, roth1970, terao1967, laves1964, waring1968, plies1980, lagergren1952, reisman1956, stephenson1971b, makovec2006, mertin1970, brennecka2007, liu2007}. 
Whereas the L-Ta$_2$O$_5$ phase is broadly described by the ``$\lambda$'' model \cite{lee2013, guo2014, hur2019, wu2020}, the precise atomic structure of H-Ta$_2$O$_5$ remains unresolved. 

The high-temperature phase can be stabilized at lower temperatures through rapid quenching or the use of oxide stabilizers, and is reported to exhibit tetragonal \cite{zaslavskii1955, roth1970, laves1964, terao1967, waring1968, plies1980} or pseudo-tetragonal \cite{lagergren1952, reisman1956, laves1964, waring1968, stephenson1971b, plies1980, makovec2006, mertin1970, brennecka2007} symmetry, both with similar lattice constants ($a \approx b \approx 3.8$~Å, $c \approx 36$~Å). Temperature-dependent studies indicate that pseudo-tetragonal forms evolve toward a tetragonal structure upon heating \cite{laves1964, waring1968, plies1980}, suggesting that the tetragonal phase represents the intrinsic high-temperature structure. 
A notable advance by Liu \textit{et al.} \cite{liu2007} achieved room-temperature stabilization of tetragonal H-Ta$_2$O$_5$ via laser irradiation without additives, resolving the tantalum sublattice in \emph{I$4_1$/amd} symmetry using transmission electron microscopy (TEM). However, the oxygen sublattice remained unresolved, leaving the full atomic structure undetermined. Density functional theory (DFT) studies \cite{sahu2004, andreoni2010, wu2011, perevalov2013, lee2013, lee2014, kim2014, perez-walton2016, yang2018, yuan2019, tong2023, he2023} have largely focused on L-Ta$_2$O$_5$, and only one stable tetragonal model has been reported via evolutionary search methods \cite{yang2018, yuan2019}. This model yields a $c$-axis lattice constant of 26.3~Å, substantially shorter than the experimental value of $\sim$36~Å, highlighting the challenge of definitively determining the crystal structure of H-Ta$_2$O$_5$.

In this study, we propose a structural model for H-Ta$_2$O$_5$ and identify an unconventional oxygen-ion migration mechanism within its stoichiometric lattice. Based on DFT calculations, we find that H-Ta$_2$O$_5$ adopts a chiral tetragonal framework constructed from orthorhombic \(\lambda\) units, forming a 168-atom unit cell. Systematic variation of the $c$-axis lattice constant shows that this structure is most stable at the experimentally reported value. Ab initio molecular dynamics simulations of this tetragonal-$\lambda$ phase reproduce the key features of the Ta sublattice observed in TEM images, providing support for the proposed model. These simulations further reveal cooperative oxygen-ion migration along single-atom-wide channels at temperatures of a few hundred degrees Celsius. The calculated migration barrier of $\sim$0.2 eV is associated with the structural flexibility of octahedral coordination near screw-rotation planes, which enables substantial lattice relaxation and accompanying charge redistribution along the migration pathways. Taken together, these results establish a distinct mechanism for ionic transport in crystalline solids and suggest new opportunities for designing fast-ion conductors based on locally flexible coordination environments.

\section*{Results}
\subsection*{\normalsize Crystal structure of H-Ta$_\mathbf{2}$O$_\mathbf{5}$}

Understanding the high-temperature H-Ta\(_2\)O\(_5\) phase builds on the widely used structural model for the low-temperature orthorhombic phase (L-Ta\(_2\)O\(_5\)). The orthorhombic “\(\lambda\)” model \cite{lee2013} consists of two-dimensional Ta\(_2\)O\(_3\) layers stacked along the \(c\)-axis and interconnected by oxygen atoms (Fig.~1a). Within each Ta\(_2\)O\(_3\) layer, all Ta atoms share an equivalent coordination environment (Fig.~1b), each bonded to three threefold-coordinated O atoms and one twofold-coordinated O atom. 
At finite temperatures, molecular dynamics simulations indicate that Ta atoms within a Ta\(_2\)O\(_3\) layer adopt a triangular arrangement (Fig.~1c), consistent with X-ray diffraction observations \cite{lehovec1964}. This triangular sublattice allows the orthorhombic-\(\lambda\) structure to be mapped onto a triangular lattice (Fig.~1d), where the twofold-coordinated O atoms—characterized by large transverse vibrations—can be viewed as links connecting neighboring lattice sites.

Our tetragonal model for H-Ta\(_2\)O\(_5\) is guided by TEM images of the Ta sublattice \cite{liu2007}, which reveal a tetragonal arrangement with $4_1$ screw symmetry (space group \emph{I\(4_1\)/amd}) and lattice constants of 3.86 Å × 3.86 Å × 36.18 Å. In this symmetry, each 90° lattice rotation is accompanied by a $c/4$ translation. Based on these features, we embed the orthorhombic-\(\lambda\) units of L-Ta\(_2\)O\(_5\) into a tetragonal framework with 4\(_1\) screw axes to construct a ``tetragonal-\(\lambda\)'' structure (Fig.~1e). 
To accommodate this motif, we adjust the lattice constants from 7.32 Å × 3.89 Å × 6.19 Å to 7.72 Å × 3.86 Å × 6.03 Å, corresponding to a volume change of only 2\%. Two additional modifications are introduced: (1) a lateral shift of \(a/4\) along the [100] direction during each 90° screw rotation to ensure proper Ta alignment at the interface, and (2) removal of selected oxygen atoms near the screw-rotation plane to preserve stoichiometry. 
The resulting pre-relaxation tetragonal structure (Fig.~1f) has lattice constants of 7.72 Å × 7.72 Å × 36.18 Å, belongs to space group \emph{P\(4_1 2_1 2\)} (No.~92) with right-handed screw symmetry, and contains 24~Ta\(_2\)O\(_5\) formula units (168 atoms). Notably, the in-plane lattice constants of the full lattice—including the oxygen sublattice—are twice those of the Ta sublattice observed in the TEM images. An enantiomorphic left-handed structure (space group \emph{P\(4_3 2_1 2\)}, No.~96) also exists (see Supplementary Note~1 and Fig.~S1).

After full DFT relaxation of the tetragonal-\(\lambda\) model (see Methods), we obtain a crystal structure for H-Ta\(_2\)O\(_5\) (Fig.~1g; atomic coordinates are provided in Supplementary Data 1 and 2). The relaxation preserves the overall space-group symmetry while modifying local bonding near the screw-rotation planes. At these planes—where two \(\lambda\) blocks meet with a 90° twist—Ta atoms adopt a distorted octahedral coordination: among the four coplanar (equatorial) oxygen neighbors, two lie slightly above and two slightly below the Ta midplane (Fig.~1i). In contrast, Ta atoms away from the screw-rotation planes retain coordination environments similar to those in the orthorhombic-\(\lambda\) structure. 
Our calculations indicate that the relaxed tetragonal-\(\lambda\) phase is lower in energy than the orthorhombic-\(\lambda\) phase by 0.10~eV per formula unit (see Supplementary Note~2 and Table~S1). The resulting lattice constants (\(a = b = 7.6\)~Å, \(c = 35.7\)~Å) are in good agreement with experimental values for H-Ta\(_2\)O\(_5\), noting that the in-plane lattice constants are approximately twice those inferred from the Ta sublattice in TEM. 
Phonon calculations for the tetragonal-$\lambda$ model (Fig.~S2) suggest that additional symmetry-lowering distortions may arise in larger unit cells. However, such refinements are expected to involve subtle modifications of the same underlying screw-rotation framework rather than a qualitatively different structural topology (see Supplementary Note~3).

To assess the stability of the tetragonal-$\lambda$ model, we examined structural variants with different $c$-axis lengths by varying the number of Ta rows between adjacent screw-rotation planes, $N_{\rm Ta}$. Guided by TEM observations, the initial model adopts $N_{\rm Ta}=3$ (Fig.~1e). For comparison, we considered structures with $N_{\rm Ta}$ ranging from 2 to 6 (Fig.~S3). For each case, we computed the screw-rotation energy \(E_{\rm SR}\), defined as the energy cost per screw-rotation plane associated with introducing screw-rotation interfaces into the orthorhombic-\(\lambda\) structure (see Methods). 
The results show that \(E_{\rm SR}<0\) only for $N_{\rm Ta}=3$ (Fig.~1h and Table~S2), indicating that incorporation of screw-rotation planes is energetically favorable exclusively in this configuration. In contrast, models with $N_{\rm Ta}\ne3$ yield \(E_{\rm SR}>0\) and exhibit noticeable strain-induced distortions (Fig.~S3). Importantly, this energetically selected configuration ($N_{\rm Ta}=3$) coincides with the periodicity inferred from TEM observations. Taken together, the combined energetic preference and experimental consistency strongly support the tetragonal-\(\lambda\) model with $N_{\rm Ta}=3$ as a realistic structural description of H-Ta\(_2\)O\(_5\), while leaving room for further refinement.

We next examine the electronic structure of the proposed H-Ta\(_2\)O\(_5\) model. Density-of-states calculations yield a band gap of $\sim$2.4~eV for the tetragonal-\(\lambda\) phase, slightly larger than the 2.2~eV gap obtained for the orthorhombic-\(\lambda\) phase (Fig.~S4). This similarity suggests that both phases maintain comparable formal oxidation states (Ta$^{5+}$ and O$^{2-}$) and a stable electronic configuration (Supplementary Note~4) \cite{lee2013}. 
We note that standard GGA functionals (including PBEsol used here) typically underestimate band gaps. Previous studies employing hybrid functionals report values of $\sim$4.0~eV for orthorhombic Ta\(_2\)O\(_5\) \cite{lee2013, guo2014, yuan2019}, consistent with the experimental gap of 3.9~eV for L-Ta\(_2\)O\(_5\) \cite{chun2003}. Although experimental band-gap measurements for H-Ta\(_2\)O\(_5\) are not yet available, the comparable GGA values suggest that a similarly large gap may be expected.

\subsection*{\normalsize Dynamical behavior of H-Ta$_\mathbf{2}$O$_\mathbf{5}$}
 
 While the proposed tetragonal-$\lambda$ model captures the zero-temperature static structure of H-Ta\(_2\)O\(_5\), subtle differences remain relative to TEM observations. Experimentally, the Ta sublattice appears achiral (space group \emph{I\(4_1\)/amd}), exhibiting no distinction between right- and left-handed screw rotations, and shows an in-plane lattice constant of 3.86 Å. In contrast, the tetragonal-\(\lambda\) model exhibits a weak chirality in the Ta sublattice and a doubled in-plane lattice constant of $\sim$7.6 Å. As shown below, these differences are effectively averaged out by finite-temperature dynamics.
 
 Our ab initio molecular dynamics simulations (20 ps at 1000 K under NVT conditions; see Methods) show that thermal fluctuations cause the instantaneous positions of Ta atoms to average into a triangular lattice (Fig.~2a). Viewed along the [010] direction, this time-averaged lattice reproduces the TEM image with \emph{I\(4_1\)/amd} symmetry and yields an effective in-plane lattice constant that is half that of the static model. A complementary projection along the [\(1\overline{1}0\)] direction (Fig.~2b) shows similarly good agreement with experiment.  These results indicate that thermal motion effectively restores the apparent achirality and reduced in-plane lattice constant of the Ta sublattice, bringing the model into agreement with TEM observations. They also provide a possible explanation for why the crystal structure of H-Ta\(_2\)O\(_5\) has remained difficult to resolve: the oxygen sublattice possesses in-plane lattice constants that are twice those commonly reported, which primarily reflect the Ta sublattice as observed in TEM.
 
 With the Ta framework established, we next examine thermal effects on the oxygen sublattice. The same molecular dynamics simulations at 1000 K reveal the emergence of oxygen migration pathways within the crystal. While most oxygen atoms fluctuate around their equilibrium positions, those located near the center of the $\lambda$ blocks—approximately midway between adjacent screw-rotation planes—exhibit anomalous, long-range motion. In the trajectories, these atoms trace continuous, wavy paths spanning multiple lattice sites and frequently overlapping (Fig.~2c).  
 Closer inspection (Fig.~2d and Supplementary Movie~1) indicates a collective drift, in which groups of oxygen atoms move coherently along the [100] or [010] directions within a Ta$_2$O$_3$ layer. As illustrated in Fig.~2e, three neighboring oxygen ions drifting along [100] maintain nearly constant separations while moving in concert. This cooperative motion involves repeated transitions between twofold and threefold coordination environments during migration (Fig.~2d). 
 The drift persists at 500 K (Fig.~S5) but is not observed at 400 K within the 20-ps simulation window, consistent with a thermally activated process. These behaviors are reproducible across multiple simulations with different initial conditions and temperatures. For comparison, analogous simulations for orthorhombic-$\lambda$ Ta$_2$O$_5$—where oxygen atoms have similar local bonding environments (Fig.~1b)—show no comparable transport even at 1500 K (Fig.~S6). 
 This contrast indicates that the tetragonal H-Ta$_2$O$_5$ structure can support oxygen mobility beyond conventional vibrational motion.
 
\subsection*{\normalsize Origin of collective oxygen transport}

To investigate the origin of the emergent oxygen mobility, we performed nudged elastic band (NEB) calculations to compare migration pathways in the tetragonal-$\lambda$ phase of H-Ta$_2$O$_5$ and the orthorhombic-$\lambda$ phase of L-Ta$_2$O$_5$. The results indicate a substantial difference in activation energies: 0.21 eV for H-Ta$_2$O$_5$ and 1.8 eV for L-Ta$_2$O$_5$. 
This contrast suggests distinct transport mechanisms in the two phases, primarily governed by differences in the structural response of the surrounding lattice.

The orthorhombic-$\lambda$ phase exhibits behavior consistent with a conventional crystalline solid, characterized by a comparatively rigid lattice framework that resists atomic rearrangement during ion migration. As an oxygen ion moves toward a neighboring site, it passes through a narrow bottleneck between two adjacent tantalum atoms (Ta1 and Ta2 in Fig.~3c). The limited lattice flexibility leads to a transition state with pronounced Ta–O bond distortions. 
Analysis of atomic displacements indicates that the response is highly localized, with significant shifts confined to the nearest and a few second-nearest neighbors of the migrating ion (Fig.~4a,b). This localized strain is accompanied by an electrostatically unfavorable accumulation of electronic charge—approximately 0.47e on each of the Ta1 and Ta2 atoms—at the transition state (Fig.~5a), reflecting the limited ability of the lattice to redistribute the perturbation. 
Together, these factors are consistent with the relatively high migration barrier of 1.8 eV. Such behavior aligns with the conventional picture of ionic transport in crystalline materials, where migration typically proceeds via hopping between vacancy or interstitial sites with only limited accommodation by the surrounding lattice.

By contrast, the tetragonal-$\lambda$ phase exhibits a more adaptive and flexible lattice response. The oxygen migration pathway obtained from NEB calculations follows a relatively broad, flat energy landscape rather than a sharply defined barrier (Fig.~3a). During migration, the surrounding lattice relaxes to accommodate the moving oxygen ion, thereby reducing local strain. In particular, the oxygen atoms bridging the screw-rotation Ta sites (O1–O4 in Fig.~3e; see also Fig.~1i) play a key role by adjusting their vertical positions, which helps alleviate strain and charge accumulation as the ion passes. 
This local flexibility propagates into longer-range structural relaxations, with atoms up to $\sim$9 Å away—including some displaced out of the Ta$_2$O$_3$ planes—responding to the migration event (Fig.~4c,d). Similar long-range relaxations have been reported in L-Ta$_2$O$_5$ models containing oxygen vacancies \cite{lee2013, guo2014, yang2014}, and have been attributed to the relatively ionic character of Ta–O bonding \cite{guo2014}. As a result, the transition-state strain is distributed over an extended region, limiting charge accumulation; the change in electronic charge on neighboring Ta atoms remains below $\sim$0.1e (Fig.~5b). 
Overall, the flexible octahedral coordination at the screw-rotation planes enables substantial structural and electronic redistribution, which is associated with a marked reduction in the migration barrier—from 1.8 eV in the orthorhombic phase to 0.2 eV in the tetragonal phase. This reduced barrier is consistent with enhanced oxygen mobility beyond simple vibrational motion at moderate temperatures. An independent estimate of the migration barrier from molecular dynamics, obtained via the temperature dependence of the diffusion coefficient, yields a value of $\sim$0.07 eV (Fig.~S7), consistent in trend with the NEB result.

\section*{Discussion}

Our findings provide a theoretical framework for interpreting longstanding experimental observations on the ionic conductivity of Ta$_2$O$_5$. Early studies \cite{mchale1981,mchale1983,choi1989,choi1990} reported that H-Ta$_2$O$_5$ exhibits oxygen ionic conductivity one to two orders of magnitude higher than that of L-Ta$_2$O$_5$, along with pronounced anisotropy. Our results offer a consistent microscopic interpretation of these observations. Specifically, we find that mobile oxygen pathways are an intrinsic feature of the tetragonal-$\lambda$ structure associated with H-Ta$_2$O$_5$, whereas such pathways are absent in the orthorhombic-$\lambda$ structure associated with L-Ta$_2$O$_5$. The identified transport channels are confined to the [100] and [010] directions, in agreement with the reported anisotropy. While the calculated migration barrier ($\sim$0.2 eV) is lower than the experimental estimate ($\sim$0.5 eV) obtained from rapidly aged single-crystal samples \cite{choi1989, choi1990}, this difference may reflect sample-dependent effects, including aging. Further measurements on well-characterized, high-purity H-Ta$_2$O$_5$ samples would be valuable for quantitatively assessing the intrinsic migration barrier.

The collective transport mechanism identified here differs in important ways from previously reported examples of cooperative ionic motion, such as those in superionic lithium conductors \cite{xu2012, deng2015, he2017, sau2024, jun2024}. In those systems, high ionic mobility is generally associated with intrinsic structural disorder or partially occupied lattice sites. In contrast, our results indicate that oxygen transport in H-Ta$_2$O$_5$ can occur within a structurally ordered, stoichiometric framework. Rather than being mediated by pre-existing defects, this behavior appears to originate from an emergent property of the lattice, namely the structural flexibility of octahedral coordination at the screw-rotation planes. These features function as local ``structural hinges,'' enabling the lattice to dynamically accommodate migrating ions and thereby supporting low-energy migration pathways. Together, these findings extend the current understanding of fast ion conduction and suggest that significant ionic mobility may be achievable in ordered crystalline materials through the presence of locally flexible coordination environments.

In summary, our combined DFT and ab initio molecular dynamics results support a tetragonal framework for H-Ta$_2$O$_5$, composed of orthorhombic-$\lambda$ units interconnected by 4$_1$ screw-rotation planes. This model reproduces the measured lattice constants and is consistent with the Ta sublattice observed in TEM images. Within this structure, oxygen ions in the Ta$_2$O$_3$ layers exhibit cooperative, one-dimensional migration above a few hundred °C, with an estimated migration barrier of $\sim$0.2~eV associated with the structural flexibility of octahedral coordination near the screw planes. These results point to a transport mechanism that differs from the conventional defect-mediated picture in ordered crystals, suggesting that adaptable coordination environments can enable ionic motion even in stoichiometric systems at moderate temperatures. More broadly, they highlight a potential design strategy for fast-ion conductors—namely, engineering locally flexible bonding environments to reduce activation barriers without relying on extrinsic defects. Further experimental validation, particularly through high-resolution characterization of the oxygen sublattice and direct measurements of migration barriers in H-Ta$_2$O$_5$, will be important for testing these predictions.

\newpage
\section*{Methods}
\subsection*{\normalsize Density functional theory calculations}

Our density functional theory (DFT) calculations were performed using the Vienna Ab Initio Simulation Package (VASP) \cite{VASP1, VASP2}. We employed the projector-augmented wave (PAW) method \cite{PAW} and the Perdew–Burke–Ernzerhof revised for solids (PBEsol) exchange–correlation functional \cite{pbesol}. A plane-wave basis with a cutoff energy of 500 eV was used for structural optimizations (400 eV for molecular dynamics simulations). Brillouin zone integrations utilized a 4×4×1 k-point mesh for the 168-atom tetragonal unit cell, and commensurate k-meshes for other structures. The lattice constants for each structure were optimized using the variable cell optimization method in VASP, with all atoms relaxed until the forces on them were less than 0.01 eV/Å. Ab initio molecular dynamics simulations were carried out in the NVT ensemble using the Nosé thermostat method \cite{md1,md2} as implemented in VASP, with a time step of 1 fs. The Nosé mass parameter was chosen to yield a temperature oscillation period of about 80 fs, ensuring adequate sampling of phase space at the target temperature. 
The phonon band structure of the tetragonal-$\lambda$ model was calculated using density-functional perturbation theory as implemented in VASP, combined with the \textsc{phonopy} code \cite{phonopy}, for the 168-atom unit cell. The oxygen diffusion coefficient $D$ was determined from the mean-square displacement (MSD) of the migrating oxygen atoms, obtained from 20~ps NVT ab initio simulations using the unit cell.
Visualization of atomic structures and trajectories was performed using VESTA software \cite{vesta}. 

\subsection*{\normalsize Screw-rotation energy calculation}

To quantify the energetic cost of introducing screw-rotation planes into the orthorhombic-\(\lambda\) structure, thereby transforming it into the tetragonal-\(\lambda\) structure, we define the screw-rotation energy $E_{\rm SR}$ per screw plane as:
\begin{equation*}
E_{\text{SR}} = \frac{1}{4} \left( E_{\text{tot}}^{\text{tetra}} - \frac{N_{\text{f.u.}}^{\text{tetra}}}{N_{\text{f.u.}}^{\text{ortho}}} E_{\text{tot}}^{\text{ortho}} \right),
\end{equation*}
where $E_{\text{tot}}^{\text{tetra}}$ and $E_{\text{tot}}^{\text{ortho}}$ are the DFT total energies (per unit cell) of the tetragonal-$\lambda$ and orthorhombic-$\lambda$ structures, respectively, and $N_{\text{f.u.}}^{\text{tetra}}$ and $N_{\text{f.u.}}^{\text{ortho}}$ are the number of Ta$_2$O$_5$ formula units in each unit cell. The factor $1/4$ accounts for four screw planes per tetragonal unit cell. A negative $E_{\rm SR}$ indicates that the tetragonal structure (with screw planes) is energetically preferred over the orthorhombic structure. For the tetragonal-\(\lambda\) structure ($N_{\rm Ta}=3$), we found $E_{\rm SR} = -0.59$ eV, reflecting a strong energetic driving force for forming the screw-rotation structure. By contrast, for the hypothetical variants with $N_{\rm Ta} \ne 3$, $E_{\rm SR}$ was positive (see Fig.~1h and Table~S2), indicating an energy penalty in those cases.

\subsection*{\normalsize Nudged elastic band (NEB) calculations}

We used the climbing-image nudged elastic band method \cite{neb1,neb2} to compute minimum energy paths and activation barriers for oxygen migration. For each path, 4–6 intermediate images were interpolated between the initial and final states, to ensure consistency. All images were relaxed until the force on each image (perpendicular to the path) was below 0.02 eV/Å. For the migration path in the tetragonal-\(\lambda\) structure, we employed the full 168-atom tetragonal unit cell with dimensions of 7.72 Å × 7.72 Å × 36.18 Å, considering three oxygen atoms moving along [100] within one Ta$_2$O$_3$ layer (the two Ta$_2$O$_3$ layers in the cell are inequivalent, but yielded nearly identical barriers, differing by $\sim$0.01~eV). To capture an equivalent migration in the orthorhombic-\(\lambda\) structure, we constructed a supercell containing two Ta$_2$O$_3$ layers (with in-plane dimensions of 7.33 Å × 24.77 Å, corresponding to a 1 × 4 repetition of the in-plane unit cell). Three oxygen atoms were moved along the shorter in-plane axis in one of the layers to simulate the migration analogous to the tetragonal case. This setup ensured that the migrating oxygen in the orthorhombic-\(\lambda\) structure had the same initial and final local environment as in the tetragonal-\(\lambda\) structure.

\newpage

\begin{thebibliography}{65}%
	\makeatletter
	\providecommand \@ifxundefined [1]{%
		\@ifx{#1\undefined}
	}%
	\providecommand \@ifnum [1]{%
		\ifnum #1\expandafter \@firstoftwo
		\else \expandafter \@secondoftwo
		\fi
	}%
	\providecommand \@ifx [1]{%
		\ifx #1\expandafter \@firstoftwo
		\else \expandafter \@secondoftwo
		\fi
	}%
	\providecommand \natexlab [1]{#1}%
	\providecommand \enquote  [1]{``#1''}%
	\providecommand \bibnamefont  [1]{#1}%
	\providecommand \bibfnamefont [1]{#1}%
	\providecommand \citenamefont [1]{#1}%
	\providecommand \href@noop [0]{\@secondoftwo}%
	\providecommand \href [0]{\begingroup \@sanitize@url \@href}%
	\providecommand \@href[1]{\@@startlink{#1}\@@href}%
	\providecommand \@@href[1]{\endgroup#1\@@endlink}%
	\providecommand \@sanitize@url [0]{\catcode `\\12\catcode `\$12\catcode
		`\&12\catcode `\#12\catcode `\^12\catcode `\_12\catcode `\%12\relax}%
	\providecommand \@@startlink[1]{}%
	\providecommand \@@endlink[0]{}%
	\providecommand \url  [0]{\begingroup\@sanitize@url \@url }%
	\providecommand \@url [1]{\endgroup\@href {#1}{\urlprefix }}%
	\providecommand \urlprefix  [0]{URL }%
	\providecommand \Eprint [0]{\href }%
	\providecommand \doibase [0]{https://doi.org/}%
	\providecommand \selectlanguage [0]{\@gobble}%
	\providecommand \bibinfo  [0]{\@secondoftwo}%
	\providecommand \bibfield  [0]{\@secondoftwo}%
	\providecommand \translation [1]{[#1]}%
	\providecommand \BibitemOpen [0]{}%
	\providecommand \bibitemStop [0]{}%
	\providecommand \bibitemNoStop [0]{.\EOS\space}%
	\providecommand \EOS [0]{\spacefactor3000\relax}%
	\providecommand \BibitemShut  [1]{\csname bibitem#1\endcsname}%
	\let\auto@bib@innerbib\@empty
	\bibitem [{\citenamefont {Farrington}\ and\ \citenamefont
		{Briant}(1979)}]{farrington1979}%
	\BibitemOpen
	\bibfield  {author} {\bibinfo {author} {\bibfnamefont {G.~C.}\ \bibnamefont
			{Farrington}}\ and\ \bibinfo {author} {\bibfnamefont {J.~L.}\ \bibnamefont
			{Briant}},\ }\bibfield  {title} {\bibinfo {title} {Fast ionic transport in
			solids},\ }\href {https://doi.org/10.1126/science.204.4400.1371} {\bibfield
		{journal} {\bibinfo  {journal} {Science}\ }\textbf {\bibinfo {volume}
			{204}},\ \bibinfo {pages} {1371} (\bibinfo {year} {1979})}\BibitemShut
	{NoStop}%
	\bibitem [{\citenamefont {Knauth}\ and\ \citenamefont
		{Tuller}(2002)}]{knauth2002}%
	\BibitemOpen
	\bibfield  {author} {\bibinfo {author} {\bibfnamefont {P.}~\bibnamefont
			{Knauth}}\ and\ \bibinfo {author} {\bibfnamefont {H.~L.}\ \bibnamefont
			{Tuller}},\ }\bibfield  {title} {\bibinfo {title} {Solid-state ionics: Roots,
			status, and future prospects},\ }\href
	{https://doi.org/https://doi.org/10.1111/j.1151-2916.2002.tb00334.x}
	{\bibfield  {journal} {\bibinfo  {journal} {J. Am. Ceram. Soc.}\ }\textbf
		{\bibinfo {volume} {85}},\ \bibinfo {pages} {1654} (\bibinfo {year}
		{2002})}\BibitemShut {NoStop}%
	\bibitem [{\citenamefont {Sau}\ \emph {et~al.}(2024)\citenamefont {Sau},
		\citenamefont {Takagi}, \citenamefont {Ikeshoji}, \citenamefont {Kisu},
		\citenamefont {Sato}, \citenamefont {dos Santos}, \citenamefont {Li},
		\citenamefont {Mohtadi},\ and\ \citenamefont {Orimo}}]{sau2024}%
	\BibitemOpen
	\bibfield  {author} {\bibinfo {author} {\bibfnamefont {K.}~\bibnamefont
			{Sau}}, \bibinfo {author} {\bibfnamefont {S.}~\bibnamefont {Takagi}},
		\bibinfo {author} {\bibfnamefont {T.}~\bibnamefont {Ikeshoji}}, \bibinfo
		{author} {\bibfnamefont {K.}~\bibnamefont {Kisu}}, \bibinfo {author}
		{\bibfnamefont {R.}~\bibnamefont {Sato}}, \bibinfo {author} {\bibfnamefont
			{E.~C.}\ \bibnamefont {dos Santos}}, \bibinfo {author} {\bibfnamefont
			{H.}~\bibnamefont {Li}}, \bibinfo {author} {\bibfnamefont {R.}~\bibnamefont
			{Mohtadi}},\ and\ \bibinfo {author} {\bibfnamefont {S.-i.}\ \bibnamefont
			{Orimo}},\ }\bibfield  {title} {\bibinfo {title} {Unlocking the secrets of
			ideal fast ion conductors for all-solid-state batteries},\ }\href
	{https://doi.org/10.1038/s43246-024-00550-z} {\bibfield  {journal} {\bibinfo
			{journal} {Commun. Mater.}\ }\textbf {\bibinfo {volume} {5}},\ \bibinfo
		{pages} {122} (\bibinfo {year} {2024})}\BibitemShut {NoStop}%
	\bibitem [{\citenamefont {Jun}\ \emph {et~al.}(2024)\citenamefont {Jun},
		\citenamefont {Chen}, \citenamefont {Wei}, \citenamefont {Yang},\ and\
		\citenamefont {Ceder}}]{jun2024}%
	\BibitemOpen
	\bibfield  {author} {\bibinfo {author} {\bibfnamefont {K.}~\bibnamefont
			{Jun}}, \bibinfo {author} {\bibfnamefont {Y.}~\bibnamefont {Chen}}, \bibinfo
		{author} {\bibfnamefont {G.}~\bibnamefont {Wei}}, \bibinfo {author}
		{\bibfnamefont {X.}~\bibnamefont {Yang}},\ and\ \bibinfo {author}
		{\bibfnamefont {G.}~\bibnamefont {Ceder}},\ }\bibfield  {title} {\bibinfo
		{title} {Diffusion mechanisms of fast lithium-ion conductors},\ }\href
	{https://doi.org/10.1038/s41578-024-00715-9} {\bibfield  {journal} {\bibinfo
			{journal} {Nat. Rev. Mater.}\ }\textbf {\bibinfo {volume} {9}},\ \bibinfo
		{pages} {887} (\bibinfo {year} {2024})}\BibitemShut {NoStop}%
	\bibitem [{\citenamefont {Sotoudeh}\ \emph {et~al.}(2024)\citenamefont
		{Sotoudeh}, \citenamefont {Baumgart}, \citenamefont {Dillenz}, \citenamefont
		{Döhn}, \citenamefont {Forster-Tonigold}, \citenamefont {Helmbrecht},
		\citenamefont {Stottmeister},\ and\ \citenamefont {Groß}}]{sotoudeh2024}%
	\BibitemOpen
	\bibfield  {author} {\bibinfo {author} {\bibfnamefont {M.}~\bibnamefont
			{Sotoudeh}}, \bibinfo {author} {\bibfnamefont {S.}~\bibnamefont {Baumgart}},
		\bibinfo {author} {\bibfnamefont {M.}~\bibnamefont {Dillenz}}, \bibinfo
		{author} {\bibfnamefont {J.}~\bibnamefont {Döhn}}, \bibinfo {author}
		{\bibfnamefont {K.}~\bibnamefont {Forster-Tonigold}}, \bibinfo {author}
		{\bibfnamefont {K.}~\bibnamefont {Helmbrecht}}, \bibinfo {author}
		{\bibfnamefont {D.}~\bibnamefont {Stottmeister}},\ and\ \bibinfo {author}
		{\bibfnamefont {A.}~\bibnamefont {Groß}},\ }\bibfield  {title} {\bibinfo
		{title} {Ion mobility in crystalline battery materials},\ }\href
	{https://doi.org/https://doi.org/10.1002/aenm.202302550} {\bibfield
		{journal} {\bibinfo  {journal} {Adv. Energy Mater.}\ }\textbf {\bibinfo
			{volume} {14}},\ \bibinfo {pages} {2302550} (\bibinfo {year}
		{2024})}\BibitemShut {NoStop}%
	\bibitem [{\citenamefont {Jacobson}(2010)}]{jacobson2010}%
	\BibitemOpen
	\bibfield  {author} {\bibinfo {author} {\bibfnamefont {A.~J.}\ \bibnamefont
			{Jacobson}},\ }\bibfield  {title} {\bibinfo {title} {Materials for solid
			oxide fuel cells},\ }\href {https://doi.org/10.1021/cm902640j} {\bibfield
		{journal} {\bibinfo  {journal} {Chem. Mater.}\ }\textbf {\bibinfo {volume}
			{22}},\ \bibinfo {pages} {660} (\bibinfo {year} {2010})}\BibitemShut
	{NoStop}%
	\bibitem [{\citenamefont {Tuller}(2017)}]{tuller2017}%
	\BibitemOpen
	\bibfield  {author} {\bibinfo {author} {\bibfnamefont {H.}~\bibnamefont
			{Tuller}},\ }\bibinfo {title} {Ionic conduction and applications},\ in\ \href
	{https://doi.org/10.1007/978-3-319-48933-9_11} {\emph {\bibinfo {booktitle}
			{Springer Handbook of Electronic and Photonic Materials}}},\ \bibinfo
	{editor} {edited by\ \bibinfo {editor} {\bibfnamefont {S.}~\bibnamefont
			{Kasap}}\ and\ \bibinfo {editor} {\bibfnamefont {P.}~\bibnamefont {Capper}}}\
	(\bibinfo  {publisher} {Springer International Publishing},\ \bibinfo
	{address} {Cham},\ \bibinfo {year} {2017})\ Chap.~\bibinfo {chapter} {11},
	pp.\ \bibinfo {pages} {247--266}\BibitemShut {NoStop}%
	\bibitem [{\citenamefont {Dwivedi}(2020)}]{dwivedi2020}%
	\BibitemOpen
	\bibfield  {author} {\bibinfo {author} {\bibfnamefont {S.}~\bibnamefont
			{Dwivedi}},\ }\bibfield  {title} {\bibinfo {title} {Solid oxide fuel cell:
			Materials for anode, cathode and electrolyte},\ }\href
	{https://doi.org/https://doi.org/10.1016/j.ijhydene.2019.11.234} {\bibfield
		{journal} {\bibinfo  {journal} {Int. J. Hydrog. Energy}\ }\textbf {\bibinfo
			{volume} {45}},\ \bibinfo {pages} {23988} (\bibinfo {year}
		{2020})}\BibitemShut {NoStop}%
	\bibitem [{\citenamefont {Brennecka}\ and\ \citenamefont
		{Payne}(2006)}]{brennecka2006}%
	\BibitemOpen
	\bibfield  {author} {\bibinfo {author} {\bibfnamefont {G.~L.}\ \bibnamefont
			{Brennecka}}\ and\ \bibinfo {author} {\bibfnamefont {D.~A.}\ \bibnamefont
			{Payne}},\ }\bibfield  {title} {\bibinfo {title} {Preparation of dense
			{Ta$_2$O$_5$}-based ceramics by a coated powder method for enhanced
			dielectric properties},\ }\href
	{https://doi.org/https://doi.org/10.1111/j.1551-2916.2006.01063.x} {\bibfield
		{journal} {\bibinfo  {journal} {J. Am. Ceram. Soc.}\ }\textbf {\bibinfo
			{volume} {89}},\ \bibinfo {pages} {2089} (\bibinfo {year}
		{2006})}\BibitemShut {NoStop}%
	\bibitem [{\citenamefont {Robertson}\ and\ \citenamefont
		{Wallace}(2015)}]{robertson2015}%
	\BibitemOpen
	\bibfield  {author} {\bibinfo {author} {\bibfnamefont {J.}~\bibnamefont
			{Robertson}}\ and\ \bibinfo {author} {\bibfnamefont {R.~M.}\ \bibnamefont
			{Wallace}},\ }\bibfield  {title} {\bibinfo {title} {{High-K materials and
				metal gates for CMOS applications}},\ }\href
	{https://doi.org/https://doi.org/10.1016/j.mser.2014.11.001} {\bibfield
		{journal} {\bibinfo  {journal} {Mater. Sci. Eng. R Rep.}\ }\textbf {\bibinfo
			{volume} {88}},\ \bibinfo {pages} {1} (\bibinfo {year} {2015})}\BibitemShut
	{NoStop}%
	\bibitem [{\citenamefont {Wang}\ \emph {et~al.}(2018)\citenamefont {Wang},
		\citenamefont {Huang}, \citenamefont {Chi}, \citenamefont {Al-Hashimi},
		\citenamefont {Marks},\ and\ \citenamefont {Facchetti}}]{wang2018}%
	\BibitemOpen
	\bibfield  {author} {\bibinfo {author} {\bibfnamefont {B.}~\bibnamefont
			{Wang}}, \bibinfo {author} {\bibfnamefont {W.}~\bibnamefont {Huang}},
		\bibinfo {author} {\bibfnamefont {L.}~\bibnamefont {Chi}}, \bibinfo {author}
		{\bibfnamefont {M.}~\bibnamefont {Al-Hashimi}}, \bibinfo {author}
		{\bibfnamefont {T.~J.}\ \bibnamefont {Marks}},\ and\ \bibinfo {author}
		{\bibfnamefont {A.}~\bibnamefont {Facchetti}},\ }\bibfield  {title} {\bibinfo
		{title} {High-$k$ gate dielectrics for emerging flexible and stretchable
			electronics},\ }\href {https://doi.org/10.1021/acs.chemrev.8b00045}
	{\bibfield  {journal} {\bibinfo  {journal} {Chem. Rev.}\ }\textbf {\bibinfo
			{volume} {118}},\ \bibinfo {pages} {5690} (\bibinfo {year}
		{2018})}\BibitemShut {NoStop}%
	\bibitem [{\citenamefont {Lee}\ \emph {et~al.}(2011)\citenamefont {Lee},
		\citenamefont {Lee}, \citenamefont {Lee}, \citenamefont {Lee}, \citenamefont
		{Chang}, \citenamefont {Hur}, \citenamefont {Kim}, \citenamefont {Kim},
		\citenamefont {Seo}, \citenamefont {Seo}, \citenamefont {Chung},
		\citenamefont {Yoo},\ and\ \citenamefont {Kim}}]{lee2011}%
	\BibitemOpen
	\bibfield  {author} {\bibinfo {author} {\bibfnamefont {M.-J.}\ \bibnamefont
			{Lee}}, \bibinfo {author} {\bibfnamefont {C.~B.}\ \bibnamefont {Lee}},
		\bibinfo {author} {\bibfnamefont {D.}~\bibnamefont {Lee}}, \bibinfo {author}
		{\bibfnamefont {S.~R.}\ \bibnamefont {Lee}}, \bibinfo {author} {\bibfnamefont
			{M.}~\bibnamefont {Chang}}, \bibinfo {author} {\bibfnamefont {J.~H.}\
			\bibnamefont {Hur}}, \bibinfo {author} {\bibfnamefont {Y.-B.}\ \bibnamefont
			{Kim}}, \bibinfo {author} {\bibfnamefont {C.-J.}\ \bibnamefont {Kim}},
		\bibinfo {author} {\bibfnamefont {D.~H.}\ \bibnamefont {Seo}}, \bibinfo
		{author} {\bibfnamefont {S.}~\bibnamefont {Seo}}, \bibinfo {author}
		{\bibfnamefont {U.-I.}\ \bibnamefont {Chung}}, \bibinfo {author}
		{\bibfnamefont {I.-K.}\ \bibnamefont {Yoo}},\ and\ \bibinfo {author}
		{\bibfnamefont {K.}~\bibnamefont {Kim}},\ }\bibfield  {title} {\bibinfo
		{title} {A fast, high-endurance and scalable non-volatile memory device made
			from asymmetric {Ta$_2$O$_{5-x}$/TaO$_{2-x}$} bilayer structures},\ }\href
	{https://doi.org/10.1038/nmat3070} {\bibfield  {journal} {\bibinfo  {journal}
			{Nat. Mater.}\ }\textbf {\bibinfo {volume} {10}},\ \bibinfo {pages} {625}
		(\bibinfo {year} {2011})}\BibitemShut {NoStop}%
	\bibitem [{\citenamefont {Kim}\ \emph {et~al.}(2020)\citenamefont {Kim},
		\citenamefont {Son}, \citenamefont {Kim}, \citenamefont {Kim}, \citenamefont
		{Lee}, \citenamefont {Park}, \citenamefont {Kwak}, \citenamefont {Park},\
		and\ \citenamefont {Jeong}}]{kim2020}%
	\BibitemOpen
	\bibfield  {author} {\bibinfo {author} {\bibfnamefont {T.}~\bibnamefont
			{Kim}}, \bibinfo {author} {\bibfnamefont {H.}~\bibnamefont {Son}}, \bibinfo
		{author} {\bibfnamefont {I.}~\bibnamefont {Kim}}, \bibinfo {author}
		{\bibfnamefont {J.}~\bibnamefont {Kim}}, \bibinfo {author} {\bibfnamefont
			{S.}~\bibnamefont {Lee}}, \bibinfo {author} {\bibfnamefont {J.~K.}\
			\bibnamefont {Park}}, \bibinfo {author} {\bibfnamefont {J.~Y.}\ \bibnamefont
			{Kwak}}, \bibinfo {author} {\bibfnamefont {J.}~\bibnamefont {Park}},\ and\
		\bibinfo {author} {\bibfnamefont {Y.}~\bibnamefont {Jeong}},\ }\bibfield
	{title} {\bibinfo {title} {{Reversible switching mode change in
				Ta$_2$O$_5$-based resistive switching memory (ReRAM)}},\ }\href
	{https://doi.org/10.1038/s41598-020-68211-y} {\bibfield  {journal} {\bibinfo
			{journal} {Sci. Rep.}\ }\textbf {\bibinfo {volume} {10}},\ \bibinfo {pages}
		{11247} (\bibinfo {year} {2020})}\BibitemShut {NoStop}%
	\bibitem [{\citenamefont {Wang}\ \emph {et~al.}(2023)\citenamefont {Wang},
		\citenamefont {Yin}, \citenamefont {Niu}, \citenamefont {Li}, \citenamefont
		{Kim},\ and\ \citenamefont {Kim}}]{wang2023}%
	\BibitemOpen
	\bibfield  {author} {\bibinfo {author} {\bibfnamefont {W.}~\bibnamefont
			{Wang}}, \bibinfo {author} {\bibfnamefont {F.}~\bibnamefont {Yin}}, \bibinfo
		{author} {\bibfnamefont {H.}~\bibnamefont {Niu}}, \bibinfo {author}
		{\bibfnamefont {Y.}~\bibnamefont {Li}}, \bibinfo {author} {\bibfnamefont
			{E.~S.}\ \bibnamefont {Kim}},\ and\ \bibinfo {author} {\bibfnamefont {N.~Y.}\
			\bibnamefont {Kim}},\ }\bibfield  {title} {\bibinfo {title} {Tantalum
			pentoxide ({Ta$_2$O$_5$} and {Ta$_2$O$_{5-x}$})-based memristor for photonic
			in-memory computing application},\ }\href
	{https://doi.org/https://doi.org/10.1016/j.nanoen.2022.108072} {\bibfield
		{journal} {\bibinfo  {journal} {Nano Energy}\ }\textbf {\bibinfo {volume}
			{106}},\ \bibinfo {pages} {108072} (\bibinfo {year} {2023})}\BibitemShut
	{NoStop}%
	\bibitem [{\citenamefont {Gurylev}(2022)}]{gurylev2022}%
	\BibitemOpen
	\bibfield  {author} {\bibinfo {author} {\bibfnamefont {V.}~\bibnamefont
			{Gurylev}},\ }\bibfield  {title} {\bibinfo {title} {A review on the
			development and advancement of {Ta$_2$O$_5$} as a promising photocatalyst},\
	}\href {https://doi.org/10.1016/j.mtsust.2022.100131} {\bibfield  {journal}
		{\bibinfo  {journal} {Mater. Today Sustain.}\ }\textbf {\bibinfo {volume}
			{18}},\ \bibinfo {pages} {100131} (\bibinfo {year} {2022})}\BibitemShut
	{NoStop}%
	\bibitem [{\citenamefont {Zheng}\ \emph {et~al.}(2022)\citenamefont {Zheng},
		\citenamefont {Vernieres}, \citenamefont {Wang}, \citenamefont {Zhang},
		\citenamefont {Hochfilzer}, \citenamefont {Krempl}, \citenamefont {Liao},
		\citenamefont {Presel}, \citenamefont {Altantzis}, \citenamefont {Fatermans},
		\citenamefont {Scott}, \citenamefont {Secher}, \citenamefont {Moon},
		\citenamefont {Liu}, \citenamefont {Bals}, \citenamefont {{Van Aert}},
		\citenamefont {Cao}, \citenamefont {Anand}, \citenamefont {N{\o}rskov},
		\citenamefont {Kibsgaard},\ and\ \citenamefont {Chorkendorff}}]{zheng2022}%
	\BibitemOpen
	\bibfield  {author} {\bibinfo {author} {\bibfnamefont {Y.~R.}\ \bibnamefont
			{Zheng}}, \bibinfo {author} {\bibfnamefont {J.}~\bibnamefont {Vernieres}},
		\bibinfo {author} {\bibfnamefont {Z.}~\bibnamefont {Wang}}, \bibinfo {author}
		{\bibfnamefont {K.}~\bibnamefont {Zhang}}, \bibinfo {author} {\bibfnamefont
			{D.}~\bibnamefont {Hochfilzer}}, \bibinfo {author} {\bibfnamefont
			{K.}~\bibnamefont {Krempl}}, \bibinfo {author} {\bibfnamefont {T.~W.}\
			\bibnamefont {Liao}}, \bibinfo {author} {\bibfnamefont {F.}~\bibnamefont
			{Presel}}, \bibinfo {author} {\bibfnamefont {T.}~\bibnamefont {Altantzis}},
		\bibinfo {author} {\bibfnamefont {J.}~\bibnamefont {Fatermans}}, \bibinfo
		{author} {\bibfnamefont {S.~B.}\ \bibnamefont {Scott}}, \bibinfo {author}
		{\bibfnamefont {N.~M.}\ \bibnamefont {Secher}}, \bibinfo {author}
		{\bibfnamefont {C.}~\bibnamefont {Moon}}, \bibinfo {author} {\bibfnamefont
			{P.}~\bibnamefont {Liu}}, \bibinfo {author} {\bibfnamefont {S.}~\bibnamefont
			{Bals}}, \bibinfo {author} {\bibfnamefont {S.}~\bibnamefont {{Van Aert}}},
		\bibinfo {author} {\bibfnamefont {A.}~\bibnamefont {Cao}}, \bibinfo {author}
		{\bibfnamefont {M.}~\bibnamefont {Anand}}, \bibinfo {author} {\bibfnamefont
			{J.~K.}\ \bibnamefont {N{\o}rskov}}, \bibinfo {author} {\bibfnamefont
			{J.}~\bibnamefont {Kibsgaard}},\ and\ \bibinfo {author} {\bibfnamefont
			{I.}~\bibnamefont {Chorkendorff}},\ }\bibfield  {title} {\bibinfo {title}
		{{Monitoring oxygen production on mass-selected iridium–tantalum oxide
				electrocatalysts}},\ }\href {https://doi.org/10.1038/s41560-021-00948-w}
	{\bibfield  {journal} {\bibinfo  {journal} {Nat. Energy}\ }\textbf {\bibinfo
			{volume} {7}},\ \bibinfo {pages} {55} (\bibinfo {year} {2022})}\BibitemShut
	{NoStop}%
	\bibitem [{\citenamefont {Sertel}\ \emph {et~al.}(2019)\citenamefont {Sertel},
		\citenamefont {Sonmez}, \citenamefont {Cetin},\ and\ \citenamefont
		{Ozcelik}}]{sertel2019}%
	\BibitemOpen
	\bibfield  {author} {\bibinfo {author} {\bibfnamefont {T.}~\bibnamefont
			{Sertel}}, \bibinfo {author} {\bibfnamefont {N.~A.}\ \bibnamefont {Sonmez}},
		\bibinfo {author} {\bibfnamefont {S.~S.}\ \bibnamefont {Cetin}},\ and\
		\bibinfo {author} {\bibfnamefont {S.}~\bibnamefont {Ozcelik}},\ }\bibfield
	{title} {\bibinfo {title} {{Influences of annealing temperature on
				anti-reflective performance of amorphous Ta$_2$O$_5$ thin films}},\ }\href
	{https://doi.org/10.1016/j.ceramint.2018.09.237} {\bibfield  {journal}
		{\bibinfo  {journal} {Ceram. Int.}\ }\textbf {\bibinfo {volume} {45}},\
		\bibinfo {pages} {11} (\bibinfo {year} {2019})}\BibitemShut {NoStop}%
	\bibitem [{\citenamefont {Zaslavskii}\ \emph {et~al.}(1955)\citenamefont
		{Zaslavskii}, \citenamefont {Zvinchuk},\ and\ \citenamefont
		{Tutov}}]{zaslavskii1955}%
	\BibitemOpen
	\bibfield  {author} {\bibinfo {author} {\bibfnamefont {A.~I.}\ \bibnamefont
			{Zaslavskii}}, \bibinfo {author} {\bibfnamefont {R.~A.}\ \bibnamefont
			{Zvinchuk}},\ and\ \bibinfo {author} {\bibfnamefont {A.~G.}\ \bibnamefont
			{Tutov}},\ }\bibfield  {title} {\bibinfo {title} {X-ray studies of
			{Ta$_2$O$_5$} polymorphism},\ }\href@noop {} {\bibfield  {journal} {\bibinfo
			{journal} {Dokl. Akad. Nauk SSSR}\ }\textbf {\bibinfo {volume} {104}},\
		\bibinfo {pages} {409} (\bibinfo {year} {1955})}\BibitemShut {NoStop}%
	\bibitem [{\citenamefont {Roth}\ \emph {et~al.}(1970)\citenamefont {Roth},
		\citenamefont {Waring},\ and\ \citenamefont {Brower}}]{roth1970}%
	\BibitemOpen
	\bibfield  {author} {\bibinfo {author} {\bibfnamefont {R.~S.}\ \bibnamefont
			{Roth}}, \bibinfo {author} {\bibfnamefont {J.~L.}\ \bibnamefont {Waring}},\
		and\ \bibinfo {author} {\bibfnamefont {W.~S.}\ \bibnamefont {Brower}},\
	}\bibfield  {title} {\bibinfo {title} {Effect of oxide additions on the
			polymorphism of tantalum pentoxide: {II. ``Stabilization"} of the high
			temperature structure type},\ }\href {https://doi.org/10.6028/jres.074A.037}
	{\bibfield  {journal} {\bibinfo  {journal} {J. Res. Natl. Bur. Stand. A Phys.
				Chem.}\ }\textbf {\bibinfo {volume} {74A}},\ \bibinfo {pages} {477–484}
		(\bibinfo {year} {1970})}\BibitemShut {NoStop}%
	\bibitem [{\citenamefont {Terao}(1967)}]{terao1967}%
	\BibitemOpen
	\bibfield  {author} {\bibinfo {author} {\bibfnamefont {N.}~\bibnamefont
			{Terao}},\ }\bibfield  {title} {\bibinfo {title} {Structure des oxides de
			tantale},\ }\href {https://doi.org/10.1143/JJAP.6.21} {\bibfield  {journal}
		{\bibinfo  {journal} {Jpn. J. Appl. Phys.}\ }\textbf {\bibinfo {volume}
			{6}},\ \bibinfo {pages} {21} (\bibinfo {year} {1967})}\BibitemShut {NoStop}%
	\bibitem [{\citenamefont {Laves}\ and\ \citenamefont
		{Petter}(1964)}]{laves1964}%
	\BibitemOpen
	\bibfield  {author} {\bibinfo {author} {\bibfnamefont {F.}~\bibnamefont
			{Laves}}\ and\ \bibinfo {author} {\bibfnamefont {W.}~\bibnamefont {Petter}},\
	}\bibfield  {title} {\bibinfo {title} {Eine displazive umwandlung bei
			$\alpha$-{Ta$_2$O$_5$}},\ }\href@noop {} {\bibfield  {journal} {\bibinfo
			{journal} {Helv. Phys. Acta}\ }\textbf {\bibinfo {volume} {37}},\ \bibinfo
		{pages} {617} (\bibinfo {year} {1964})}\BibitemShut {NoStop}%
	\bibitem [{\citenamefont {Waring}\ and\ \citenamefont
		{Roth}(1968)}]{waring1968}%
	\BibitemOpen
	\bibfield  {author} {\bibinfo {author} {\bibfnamefont {J.~L.}\ \bibnamefont
			{Waring}}\ and\ \bibinfo {author} {\bibfnamefont {R.~S.}\ \bibnamefont
			{Roth}},\ }\bibfield  {title} {\bibinfo {title} {Effect of oxide additions on
			polymorphism of tantalum pentoxide (system {Ta$_2$O$_5$-TiO$_2$)}},\ }\href
	{https://doi.org/10.6028/jres.072a.018} {\bibfield  {journal} {\bibinfo
			{journal} {J. Res. Natl. Bur. Stand. A Phys. Chem.}\ }\textbf {\bibinfo
			{volume} {72A}},\ \bibinfo {pages} {175} (\bibinfo {year}
		{1968})}\BibitemShut {NoStop}%
	\bibitem [{\citenamefont {Plies}\ and\ \citenamefont
		{Gruehn}(1980)}]{plies1980}%
	\BibitemOpen
	\bibfield  {author} {\bibinfo {author} {\bibfnamefont {V.~V.}\ \bibnamefont
			{Plies}}\ and\ \bibinfo {author} {\bibfnamefont {R.}~\bibnamefont {Gruehn}},\
	}\bibfield  {title} {\bibinfo {title} {Zum thermischen verhalten metastabiler
			{H-Ta$_2$O$_5$}-varianten im system {Ta$_2$O$_5$-TiO$_2$}},\ }\href
	{https://doi.org/https://doi.org/10.1002/zaac.19804630105} {\bibfield
		{journal} {\bibinfo  {journal} {Z. Anorg. Allg. Chem.}\ }\textbf {\bibinfo
			{volume} {463}},\ \bibinfo {pages} {32} (\bibinfo {year} {1980})}\BibitemShut
	{NoStop}%
	\bibitem [{\citenamefont {Lagergren}\ and\ \citenamefont
		{Magn{\'{e}}li}(1952)}]{lagergren1952}%
	\BibitemOpen
	\bibfield  {author} {\bibinfo {author} {\bibfnamefont {S.}~\bibnamefont
			{Lagergren}}\ and\ \bibinfo {author} {\bibfnamefont {A.}~\bibnamefont
			{Magn{\'{e}}li}},\ }\bibfield  {title} {\bibinfo {title} {On the
			tantalum-oxygen system},\ }\href
	{https://doi.org/10.3891/acta.chem.scand.06-0444} {\bibfield  {journal}
		{\bibinfo  {journal} {Acta Chem. Scand.}\ }\textbf {\bibinfo {volume} {6}},\
		\bibinfo {pages} {444} (\bibinfo {year} {1952})},\ \bibinfo {note} {and ASTM
		Card 5-0258}\BibitemShut {NoStop}%
	\bibitem [{\citenamefont {Reisman}\ \emph {et~al.}(1956)\citenamefont
		{Reisman}, \citenamefont {Holtzberg}, \citenamefont {Berkenblit},\ and\
		\citenamefont {Berry}}]{reisman1956}%
	\BibitemOpen
	\bibfield  {author} {\bibinfo {author} {\bibfnamefont {A.}~\bibnamefont
			{Reisman}}, \bibinfo {author} {\bibfnamefont {F.}~\bibnamefont {Holtzberg}},
		\bibinfo {author} {\bibfnamefont {M.}~\bibnamefont {Berkenblit}},\ and\
		\bibinfo {author} {\bibfnamefont {M.}~\bibnamefont {Berry}},\ }\bibfield
	{title} {\bibinfo {title} {Reactions of the group {VB} pentoxides with alkali
			oxides and carbonates. {III}. thermal and {X}-ray phase diagrams of the
			system {K$_2$O} or {K$_2$CO$_3$} with {Ta$_2$O$_5$}},\ }\href
	{https://doi.org/10.1021/ja01599a003} {\bibfield  {journal} {\bibinfo
			{journal} {J. Am. Chem. Soc.}\ }\textbf {\bibinfo {volume} {78}},\ \bibinfo
		{pages} {4514} (\bibinfo {year} {1956})}\BibitemShut {NoStop}%
	\bibitem [{\citenamefont {Stephenson}\ and\ \citenamefont
		{Roth}(1971)}]{stephenson1971b}%
	\BibitemOpen
	\bibfield  {author} {\bibinfo {author} {\bibfnamefont {N.}~\bibnamefont
			{Stephenson}}\ and\ \bibinfo {author} {\bibfnamefont {R.}~\bibnamefont
			{Roth}},\ }\bibfield  {title} {\bibinfo {title} {The crystal structure of the
			high temperature form of {Ta$_2$O$_5$}},\ }\href
	{https://doi.org/https://doi.org/10.1016/0022-4596(71)90018-1} {\bibfield
		{journal} {\bibinfo  {journal} {J. Solid State Chem.}\ }\textbf {\bibinfo
			{volume} {3}},\ \bibinfo {pages} {145} (\bibinfo {year} {1971})}\BibitemShut
	{NoStop}%
	\bibitem [{\citenamefont {Makovec}\ \emph {et~al.}(2006)\citenamefont
		{Makovec}, \citenamefont {Zuo}, \citenamefont {Twesten},\ and\ \citenamefont
		{Payne}}]{makovec2006}%
	\BibitemOpen
	\bibfield  {author} {\bibinfo {author} {\bibfnamefont {D.}~\bibnamefont
			{Makovec}}, \bibinfo {author} {\bibfnamefont {J.-M.}\ \bibnamefont {Zuo}},
		\bibinfo {author} {\bibfnamefont {R.}~\bibnamefont {Twesten}},\ and\ \bibinfo
		{author} {\bibfnamefont {D.~A.}\ \bibnamefont {Payne}},\ }\bibfield  {title}
	{\bibinfo {title} {A high-temperature structure for {Ta$_2$O$_5$} with
			modulations by {TiO$_2$} substitution},\ }\href
	{https://doi.org/https://doi.org/10.1016/j.jssc.2006.03.014} {\bibfield
		{journal} {\bibinfo  {journal} {J. Solid State Chem.}\ }\textbf {\bibinfo
			{volume} {179}},\ \bibinfo {pages} {1782} (\bibinfo {year}
		{2006})}\BibitemShut {NoStop}%
	\bibitem [{\citenamefont {Mertin}\ \emph {et~al.}(1970)\citenamefont {Mertin},
		\citenamefont {Gruehn},\ and\ \citenamefont {Schäfer}}]{mertin1970}%
	\BibitemOpen
	\bibfield  {author} {\bibinfo {author} {\bibfnamefont {W.}~\bibnamefont
			{Mertin}}, \bibinfo {author} {\bibfnamefont {R.}~\bibnamefont {Gruehn}},\
		and\ \bibinfo {author} {\bibfnamefont {H.}~\bibnamefont {Schäfer}},\
	}\bibfield  {title} {\bibinfo {title} {Neue beobachtungen zum system
			{TiO$_2$-Ta$_2$O$_5$}},\ }\href
	{https://doi.org/https://doi.org/10.1016/0022-4596(70)90125-8} {\bibfield
		{journal} {\bibinfo  {journal} {J. Solid State Chem.}\ }\textbf {\bibinfo
			{volume} {1}},\ \bibinfo {pages} {425} (\bibinfo {year} {1970})}\BibitemShut
	{NoStop}%
	\bibitem [{\citenamefont {Brennecka}\ \emph {et~al.}(2007)\citenamefont
		{Brennecka}, \citenamefont {Payne}, \citenamefont {Sarin}, \citenamefont
		{Zuo}, \citenamefont {Kriven},\ and\ \citenamefont
		{Hellwig}}]{brennecka2007}%
	\BibitemOpen
	\bibfield  {author} {\bibinfo {author} {\bibfnamefont {G.~L.}\ \bibnamefont
			{Brennecka}}, \bibinfo {author} {\bibfnamefont {D.~A.}\ \bibnamefont
			{Payne}}, \bibinfo {author} {\bibfnamefont {P.}~\bibnamefont {Sarin}},
		\bibinfo {author} {\bibfnamefont {J.-M.}\ \bibnamefont {Zuo}}, \bibinfo
		{author} {\bibfnamefont {W.~M.}\ \bibnamefont {Kriven}},\ and\ \bibinfo
		{author} {\bibfnamefont {H.}~\bibnamefont {Hellwig}},\ }\bibfield  {title}
	{\bibinfo {title} {Phase transformations in the high-temperature form of pure
			and {TiO$_2$}-stabilized {Ta$_2$O$_5$}},\ }\href
	{https://doi.org/https://doi.org/10.1111/j.1551-2916.2007.01790.x} {\bibfield
		{journal} {\bibinfo  {journal} {J. Am. Ceram. Soc.}\ }\textbf {\bibinfo
			{volume} {90}},\ \bibinfo {pages} {2947} (\bibinfo {year}
		{2007})}\BibitemShut {NoStop}%
	\bibitem [{\citenamefont {Liu}\ \emph {et~al.}(2007)\citenamefont {Liu},
		\citenamefont {Han}, \citenamefont {Zhang}, \citenamefont {Ji},\ and\
		\citenamefont {Jiang}}]{liu2007}%
	\BibitemOpen
	\bibfield  {author} {\bibinfo {author} {\bibfnamefont {X.~Q.}\ \bibnamefont
			{Liu}}, \bibinfo {author} {\bibfnamefont {X.~D.}\ \bibnamefont {Han}},
		\bibinfo {author} {\bibfnamefont {Z.}~\bibnamefont {Zhang}}, \bibinfo
		{author} {\bibfnamefont {L.~F.}\ \bibnamefont {Ji}},\ and\ \bibinfo {author}
		{\bibfnamefont {Y.~J.}\ \bibnamefont {Jiang}},\ }\bibfield  {title} {\bibinfo
		{title} {{The crystal structure of high temperature phase Ta$_2$O$_5$}},\
	}\href {https://doi.org/10.1016/j.actamat.2006.11.031} {\bibfield  {journal}
		{\bibinfo  {journal} {Acta Mater.}\ }\textbf {\bibinfo {volume} {55}},\
		\bibinfo {pages} {2385} (\bibinfo {year} {2007})}\BibitemShut {NoStop}%
	\bibitem [{\citenamefont {Lee}\ \emph {et~al.}(2013)\citenamefont {Lee},
		\citenamefont {Kim}, \citenamefont {Kim}, \citenamefont {Kim},\ and\
		\citenamefont {Park}}]{lee2013}%
	\BibitemOpen
	\bibfield  {author} {\bibinfo {author} {\bibfnamefont {S.-H.}\ \bibnamefont
			{Lee}}, \bibinfo {author} {\bibfnamefont {J.}~\bibnamefont {Kim}}, \bibinfo
		{author} {\bibfnamefont {S.-J.}\ \bibnamefont {Kim}}, \bibinfo {author}
		{\bibfnamefont {S.}~\bibnamefont {Kim}},\ and\ \bibinfo {author}
		{\bibfnamefont {G.-S.}\ \bibnamefont {Park}},\ }\bibfield  {title} {\bibinfo
		{title} {Hidden structural order in orthorhombic {Ta$_2$O$_5$}},\ }\href
	{https://doi.org/10.1103/PhysRevLett.110.235502} {\bibfield  {journal}
		{\bibinfo  {journal} {Phys. Rev. Lett.}\ }\textbf {\bibinfo {volume} {110}},\
		\bibinfo {pages} {235502} (\bibinfo {year} {2013})}\BibitemShut {NoStop}%
	\bibitem [{\citenamefont {Guo}\ and\ \citenamefont
		{Robertson}(2014)}]{guo2014}%
	\BibitemOpen
	\bibfield  {author} {\bibinfo {author} {\bibfnamefont {Y.}~\bibnamefont
			{Guo}}\ and\ \bibinfo {author} {\bibfnamefont {J.}~\bibnamefont
			{Robertson}},\ }\bibfield  {title} {\bibinfo {title} {{Oxygen vacancy defects
				in Ta$_2$O$_5$ showing long-range atomic re-arrangements}},\ }\href
	{https://doi.org/10.1063/1.4869553} {\bibfield  {journal} {\bibinfo
			{journal} {Appl. Phys. Lett.}\ }\textbf {\bibinfo {volume} {104}},\ \bibinfo
		{pages} {112906} (\bibinfo {year} {2014})}\BibitemShut {NoStop}%
	\bibitem [{\citenamefont {Hur}(2019)}]{hur2019}%
	\BibitemOpen
	\bibfield  {author} {\bibinfo {author} {\bibfnamefont {J.~H.}\ \bibnamefont
			{Hur}},\ }\bibfield  {title} {\bibinfo {title} {{Theoretical studies on
				oxygen vacancy migration energy barrier in the orthorhombic $\lambda$ phase
				Ta$_2$O$_5$}},\ }\href {https://doi.org/10.1016/j.commatsci.2019.109148}
	{\bibfield  {journal} {\bibinfo  {journal} {Comput. Mater. Sci.}\ }\textbf
		{\bibinfo {volume} {169}},\ \bibinfo {pages} {109148} (\bibinfo {year}
		{2019})}\BibitemShut {NoStop}%
	\bibitem [{\citenamefont {Wu}\ \emph {et~al.}(2020)\citenamefont {Wu},
		\citenamefont {Cui}, \citenamefont {Zhang}, \citenamefont {Wang},
		\citenamefont {Huang}, \citenamefont {Zhang},\ and\ \citenamefont
		{Xu}}]{wu2020}%
	\BibitemOpen
	\bibfield  {author} {\bibinfo {author} {\bibfnamefont {X.}~\bibnamefont
			{Wu}}, \bibinfo {author} {\bibfnamefont {N.}~\bibnamefont {Cui}}, \bibinfo
		{author} {\bibfnamefont {Q.}~\bibnamefont {Zhang}}, \bibinfo {author}
		{\bibfnamefont {W.}~\bibnamefont {Wang}}, \bibinfo {author} {\bibfnamefont
			{Q.}~\bibnamefont {Huang}}, \bibinfo {author} {\bibfnamefont
			{H.}~\bibnamefont {Zhang}},\ and\ \bibinfo {author} {\bibfnamefont
			{Q.}~\bibnamefont {Xu}},\ }\bibfield  {title} {\bibinfo {title} {{Density
				functional theory study of oxygen vacancy defect diffusion properties in
				$\lambda$-Ta$_2$O$_5$}},\ }\href {https://doi.org/10.35848/1347-4065/abcc14}
	{\bibfield  {journal} {\bibinfo  {journal} {Jap. J. Appl. Phys.}\ }\textbf
		{\bibinfo {volume} {59}},\ \bibinfo {pages} {121003} (\bibinfo {year}
		{2020})}\BibitemShut {NoStop}%
	\bibitem [{\citenamefont {Sahu}\ and\ \citenamefont
		{Kleinman}(2004)}]{sahu2004}%
	\BibitemOpen
	\bibfield  {author} {\bibinfo {author} {\bibfnamefont {B.~R.}\ \bibnamefont
			{Sahu}}\ and\ \bibinfo {author} {\bibfnamefont {L.}~\bibnamefont
			{Kleinman}},\ }\bibfield  {title} {\bibinfo {title} {{Theoretical study of
				structural and electronic properties of $\beta$-Ta$_2$O$_5$ and
				$\delta$-Ta$_2$O$_5$}},\ }\href {https://doi.org/10.1103/PhysRevB.69.165202}
	{\bibfield  {journal} {\bibinfo  {journal} {Phys. Rev. B}\ }\textbf {\bibinfo
			{volume} {69}},\ \bibinfo {pages} {1} (\bibinfo {year} {2004})}\BibitemShut
	{NoStop}%
	\bibitem [{\citenamefont {Andreoni}\ and\ \citenamefont
		{Pignedoli}(2010)}]{andreoni2010}%
	\BibitemOpen
	\bibfield  {author} {\bibinfo {author} {\bibfnamefont {W.}~\bibnamefont
			{Andreoni}}\ and\ \bibinfo {author} {\bibfnamefont {C.~A.}\ \bibnamefont
			{Pignedoli}},\ }\bibfield  {title} {\bibinfo {title} {{Ta$_2$O$_5$}
			polymorphs: Structural motifs and dielectric constant from first
			principles},\ }\href {https://doi.org/10.1063/1.3308475} {\bibfield
		{journal} {\bibinfo  {journal} {Appl. Phys. Lett.}\ }\textbf {\bibinfo
			{volume} {96}},\ \bibinfo {pages} {2010} (\bibinfo {year}
		{2010})}\BibitemShut {NoStop}%
	\bibitem [{\citenamefont {Wu}\ \emph {et~al.}(2011)\citenamefont {Wu},
		\citenamefont {Li},\ and\ \citenamefont {Cheng}}]{wu2011}%
	\BibitemOpen
	\bibfield  {author} {\bibinfo {author} {\bibfnamefont {Y.~N.}\ \bibnamefont
			{Wu}}, \bibinfo {author} {\bibfnamefont {L.}~\bibnamefont {Li}},\ and\
		\bibinfo {author} {\bibfnamefont {H.~P.}\ \bibnamefont {Cheng}},\ }\bibfield
	{title} {\bibinfo {title} {{First-principles studies of Ta$_2$O$_5$
				polymorphs}},\ }\href {https://doi.org/10.1103/PhysRevB.83.144105} {\bibfield
		{journal} {\bibinfo  {journal} {Phys. Rev. B}\ }\textbf {\bibinfo {volume}
			{83}},\ \bibinfo {pages} {1} (\bibinfo {year} {2011})}\BibitemShut {NoStop}%
	\bibitem [{\citenamefont {Perevalov}\ and\ \citenamefont
		{Shaposhnikov}(2013)}]{perevalov2013}%
	\BibitemOpen
	\bibfield  {author} {\bibinfo {author} {\bibfnamefont {T.~V.}\ \bibnamefont
			{Perevalov}}\ and\ \bibinfo {author} {\bibfnamefont {A.~V.}\ \bibnamefont
			{Shaposhnikov}},\ }\bibfield  {title} {\bibinfo {title} {{Ab initio
				simulation of the electronic structure of Ta$_2$O$_5$ crystal
				modifications}},\ }\href {https://doi.org/10.1134/S1063776113040158}
	{\bibfield  {journal} {\bibinfo  {journal} {J. Exp. Theor. Phys.}\ }\textbf
		{\bibinfo {volume} {116}},\ \bibinfo {pages} {995} (\bibinfo {year}
		{2013})}\BibitemShut {NoStop}%
	\bibitem [{\citenamefont {Lee}\ \emph {et~al.}(2014)\citenamefont {Lee},
		\citenamefont {Lu},\ and\ \citenamefont {Kioupakis}}]{lee2014}%
	\BibitemOpen
	\bibfield  {author} {\bibinfo {author} {\bibfnamefont {J.}~\bibnamefont
			{Lee}}, \bibinfo {author} {\bibfnamefont {W.}~\bibnamefont {Lu}},\ and\
		\bibinfo {author} {\bibfnamefont {E.}~\bibnamefont {Kioupakis}},\ }\bibfield
	{title} {\bibinfo {title} {{Electronic properties of tantalum pentoxide
				polymorphs from first-principles calculations}},\ }\href
	{https://doi.org/10.1063/1.4901939} {\bibfield  {journal} {\bibinfo
			{journal} {Appl. Phys. Lett.}\ }\textbf {\bibinfo {volume} {105}},\ \bibinfo
		{pages} {202108} (\bibinfo {year} {2014})}\BibitemShut {NoStop}%
	\bibitem [{\citenamefont {Kim}\ \emph {et~al.}(2014)\citenamefont {Kim},
		\citenamefont {Magyari-Köpe}, \citenamefont {Lee}, \citenamefont {Kim},
		\citenamefont {Lee},\ and\ \citenamefont {Nishi}}]{kim2014}%
	\BibitemOpen
	\bibfield  {author} {\bibinfo {author} {\bibfnamefont {J.-Y.}\ \bibnamefont
			{Kim}}, \bibinfo {author} {\bibfnamefont {B.}~\bibnamefont {Magyari-Köpe}},
		\bibinfo {author} {\bibfnamefont {K.-J.}\ \bibnamefont {Lee}}, \bibinfo
		{author} {\bibfnamefont {H.-S.}\ \bibnamefont {Kim}}, \bibinfo {author}
		{\bibfnamefont {S.-H.}\ \bibnamefont {Lee}},\ and\ \bibinfo {author}
		{\bibfnamefont {Y.}~\bibnamefont {Nishi}},\ }\bibfield  {title} {\bibinfo
		{title} {{Electronic structure and stability of low symmetry Ta$_2$O$_5$
				polymorphs}},\ }\href {https://doi.org/10.1002/pssr.201409018} {\bibfield
		{journal} {\bibinfo  {journal} {Phys. Status Solidi RRL}\ }\textbf {\bibinfo
			{volume} {8}},\ \bibinfo {pages} {560} (\bibinfo {year} {2014})}\BibitemShut
	{NoStop}%
	\bibitem [{\citenamefont {P{\'{e}}rez-Walton}\ \emph
		{et~al.}(2016)\citenamefont {P{\'{e}}rez-Walton}, \citenamefont
		{Valencia-Balv{\'{i}}n}, \citenamefont {Padilha}, \citenamefont {Dalpian},\
		and\ \citenamefont {Osorio-Guill{\'{e}}n}}]{perez-walton2016}%
	\BibitemOpen
	\bibfield  {author} {\bibinfo {author} {\bibfnamefont {S.}~\bibnamefont
			{P{\'{e}}rez-Walton}}, \bibinfo {author} {\bibfnamefont {C.}~\bibnamefont
			{Valencia-Balv{\'{i}}n}}, \bibinfo {author} {\bibfnamefont {A.~C.}\
			\bibnamefont {Padilha}}, \bibinfo {author} {\bibfnamefont {G.~M.}\
			\bibnamefont {Dalpian}},\ and\ \bibinfo {author} {\bibfnamefont {J.~M.}\
			\bibnamefont {Osorio-Guill{\'{e}}n}},\ }\bibfield  {title} {\bibinfo {title}
		{{A search for the ground state structure and the phase stability of tantalum
				pentoxide}},\ }\href {https://doi.org/10.1088/0953-8984/28/3/035801}
	{\bibfield  {journal} {\bibinfo  {journal} {J. Phys. Condens. Matter}\
		}\textbf {\bibinfo {volume} {28}},\ \bibinfo {pages} {035801} (\bibinfo
		{year} {2016})}\BibitemShut {NoStop}%
	\bibitem [{\citenamefont {Yang}\ and\ \citenamefont
		{Kawazoe}(2018)}]{yang2018}%
	\BibitemOpen
	\bibfield  {author} {\bibinfo {author} {\bibfnamefont {Y.}~\bibnamefont
			{Yang}}\ and\ \bibinfo {author} {\bibfnamefont {Y.}~\bibnamefont {Kawazoe}},\
	}\bibfield  {title} {\bibinfo {title} {Prediction of new ground-state crystal
			structure of {Ta$_2$O$_5$}},\ }\href
	{https://doi.org/10.1103/PhysRevMaterials.2.034602} {\bibfield  {journal}
		{\bibinfo  {journal} {Phys. Rev. Mater.}\ }\textbf {\bibinfo {volume} {2}},\
		\bibinfo {pages} {034602} (\bibinfo {year} {2018})}\BibitemShut {NoStop}%
	\bibitem [{\citenamefont {Yuan}\ \emph {et~al.}(2019)\citenamefont {Yuan},
		\citenamefont {Xue}, \citenamefont {Chen}, \citenamefont {Fonseca},\ and\
		\citenamefont {Miao}}]{yuan2019}%
	\BibitemOpen
	\bibfield  {author} {\bibinfo {author} {\bibfnamefont {J.-H.}\ \bibnamefont
			{Yuan}}, \bibinfo {author} {\bibfnamefont {K.-H.}\ \bibnamefont {Xue}},
		\bibinfo {author} {\bibfnamefont {Q.}~\bibnamefont {Chen}}, \bibinfo {author}
		{\bibfnamefont {L.~R.~C.}\ \bibnamefont {Fonseca}},\ and\ \bibinfo {author}
		{\bibfnamefont {X.-S.}\ \bibnamefont {Miao}},\ }\bibfield  {title} {\bibinfo
		{title} {Ab initio simulation of {Ta$_2$O$_5$}: A high symmetry ground state
			phase with application to interface calculation},\ }\href
	{https://doi.org/10.1002/andp.201800524} {\bibfield  {journal} {\bibinfo
			{journal} {Ann. Phys.}\ }\textbf {\bibinfo {volume} {531}},\ \bibinfo {pages}
		{1800524} (\bibinfo {year} {2019})}\BibitemShut {NoStop}%
	\bibitem [{\citenamefont {Tong}\ \emph {et~al.}(2023)\citenamefont {Tong},
		\citenamefont {Tang},\ and\ \citenamefont {Yang}}]{tong2023}%
	\BibitemOpen
	\bibfield  {author} {\bibinfo {author} {\bibfnamefont {Y.}~\bibnamefont
			{Tong}}, \bibinfo {author} {\bibfnamefont {H.}~\bibnamefont {Tang}},\ and\
		\bibinfo {author} {\bibfnamefont {Y.}~\bibnamefont {Yang}},\ }\bibfield
	{title} {\bibinfo {title} {{Structural and electronic properties of
				Ta$_2$O$_5$ with one formula unit}},\ }\href
	{https://doi.org/10.1016/j.commatsci.2023.112482} {\bibfield  {journal}
		{\bibinfo  {journal} {Comput. Mater. Sci.}\ }\textbf {\bibinfo {volume}
			{230}},\ \bibinfo {pages} {112482} (\bibinfo {year} {2023})}\BibitemShut
	{NoStop}%
	\bibitem [{\citenamefont {He}\ and\ \citenamefont {Sun}(2023)}]{he2023}%
	\BibitemOpen
	\bibfield  {author} {\bibinfo {author} {\bibfnamefont {Y.}~\bibnamefont
			{He}}\ and\ \bibinfo {author} {\bibfnamefont {H.}~\bibnamefont {Sun}},\
	}\bibfield  {title} {\bibinfo {title} {{Unusual mechanical strengths of
				Ta$_2$O$_5$stable phases: A first-principles calculation study}},\ }\href
	{https://doi.org/10.1063/5.0138458} {\bibfield  {journal} {\bibinfo
			{journal} {J. Appl. Phys.}\ }\textbf {\bibinfo {volume} {133}},\ \bibinfo
		{pages} {095103} (\bibinfo {year} {2023})}\BibitemShut {NoStop}%
	\bibitem [{\citenamefont {Lehovec}(1964)}]{lehovec1964}%
	\BibitemOpen
	\bibfield  {author} {\bibinfo {author} {\bibfnamefont {K.}~\bibnamefont
			{Lehovec}},\ }\bibfield  {title} {\bibinfo {title} {{Lattice structure of
				$\beta$-Ta$_2$O$_5$}},\ }\href {https://doi.org/10.1016/0022-5088(64)90036-0}
	{\bibfield  {journal} {\bibinfo  {journal} {J. Less Common Met.}\ }\textbf
		{\bibinfo {volume} {7}},\ \bibinfo {pages} {397} (\bibinfo {year}
		{1964})}\BibitemShut {NoStop}%
	\bibitem [{\citenamefont {Chun}\ \emph {et~al.}(2003)\citenamefont {Chun},
		\citenamefont {Ishikawa}, \citenamefont {Fujisawa}, \citenamefont {Takata},
		\citenamefont {Kondo}, \citenamefont {Hara}, \citenamefont {Kawai},
		\citenamefont {Matsumoto},\ and\ \citenamefont {Domen}}]{chun2003}%
	\BibitemOpen
	\bibfield  {author} {\bibinfo {author} {\bibfnamefont {W.~J.}\ \bibnamefont
			{Chun}}, \bibinfo {author} {\bibfnamefont {A.}~\bibnamefont {Ishikawa}},
		\bibinfo {author} {\bibfnamefont {H.}~\bibnamefont {Fujisawa}}, \bibinfo
		{author} {\bibfnamefont {T.}~\bibnamefont {Takata}}, \bibinfo {author}
		{\bibfnamefont {J.~N.}\ \bibnamefont {Kondo}}, \bibinfo {author}
		{\bibfnamefont {M.}~\bibnamefont {Hara}}, \bibinfo {author} {\bibfnamefont
			{M.}~\bibnamefont {Kawai}}, \bibinfo {author} {\bibfnamefont
			{Y.}~\bibnamefont {Matsumoto}},\ and\ \bibinfo {author} {\bibfnamefont
			{K.}~\bibnamefont {Domen}},\ }\bibfield  {title} {\bibinfo {title}
		{{Conduction and valence band positions of Ta$_2$O$_5$, TaON, and Ta$_3$N$_5$
				by UPS and electrochemical methods}},\ }\href
	{https://doi.org/10.1021/jp027593f} {\bibfield  {journal} {\bibinfo
			{journal} {J. Phys. Chem. B}\ }\textbf {\bibinfo {volume} {107}},\ \bibinfo
		{pages} {1798} (\bibinfo {year} {2003})}\BibitemShut {NoStop}%
	\bibitem [{\citenamefont {Yang}\ \emph {et~al.}(2014)\citenamefont {Yang},
		\citenamefont {Sugino},\ and\ \citenamefont {Kawazoe}}]{yang2014}%
	\BibitemOpen
	\bibfield  {author} {\bibinfo {author} {\bibfnamefont {Y.}~\bibnamefont
			{Yang}}, \bibinfo {author} {\bibfnamefont {O.}~\bibnamefont {Sugino}},\ and\
		\bibinfo {author} {\bibfnamefont {Y.}~\bibnamefont {Kawazoe}},\ }\bibfield
	{title} {\bibinfo {title} {{Exceptionally long-ranged lattice relaxation in
				oxygen-deficient Ta$_2$O$_5$}},\ }\href
	{https://doi.org/10.1016/j.ssc.2014.06.021} {\bibfield  {journal} {\bibinfo
			{journal} {Solid State Commun.}\ }\textbf {\bibinfo {volume} {195}},\
		\bibinfo {pages} {16} (\bibinfo {year} {2014})}\BibitemShut {NoStop}%
	\bibitem [{\citenamefont {McHale}\ and\ \citenamefont
		{Tuller}(1981)}]{mchale1981}%
	\BibitemOpen
	\bibfield  {author} {\bibinfo {author} {\bibfnamefont {A.~E.}\ \bibnamefont
			{McHale}}\ and\ \bibinfo {author} {\bibfnamefont {H.~L.}\ \bibnamefont
			{Tuller}},\ }\bibfield  {title} {\bibinfo {title} {New tantala-based solid
			oxide electrolytes},\ }\href
	{https://doi.org/https://doi.org/10.1016/0167-2738(81)90305-2} {\bibfield
		{journal} {\bibinfo  {journal} {Solid State Ionics}\ }\textbf {\bibinfo
			{volume} {5}},\ \bibinfo {pages} {515} (\bibinfo {year} {1981})}\BibitemShut
	{NoStop}%
	\bibitem [{\citenamefont {McHale}\ and\ \citenamefont
		{Tuller}(1983)}]{mchale1983}%
	\BibitemOpen
	\bibfield  {author} {\bibinfo {author} {\bibfnamefont {A.~E.}\ \bibnamefont
			{McHale}}\ and\ \bibinfo {author} {\bibfnamefont {H.~L.}\ \bibnamefont
			{Tuller}},\ }\bibfield  {title} {\bibinfo {title} {Defects and charge
			transport in stabilized $\alpha$-{Ta$_2$O$_5$}},\ }\href
	{https://doi.org/10.1080/00337578308224711} {\bibfield  {journal} {\bibinfo
			{journal} {Radia. Eff.}\ }\textbf {\bibinfo {volume} {75}},\ \bibinfo {pages}
		{267} (\bibinfo {year} {1983})}\BibitemShut {NoStop}%
	\bibitem [{\citenamefont {Choi}\ \emph {et~al.}(1989)\citenamefont {Choi},
		\citenamefont {Tuller},\ and\ \citenamefont {Haggerty}}]{choi1989}%
	\BibitemOpen
	\bibfield  {author} {\bibinfo {author} {\bibfnamefont {G.~M.}\ \bibnamefont
			{Choi}}, \bibinfo {author} {\bibfnamefont {H.~L.}\ \bibnamefont {Tuller}},\
		and\ \bibinfo {author} {\bibfnamefont {J.~S.}\ \bibnamefont {Haggerty}},\
	}\bibfield  {title} {\bibinfo {title} {Alpha-{Ta$_2$O$_5$}: An intrinsic fast
			oxygen ion conductor},\ }\href {https://doi.org/10.1149/1.2096752} {\bibfield
		{journal} {\bibinfo  {journal} {J. Electrochem. Soc.}\ }\textbf {\bibinfo
			{volume} {136}},\ \bibinfo {pages} {835} (\bibinfo {year}
		{1989})}\BibitemShut {NoStop}%
	\bibitem [{\citenamefont {Choi}\ and\ \citenamefont {Tuller}(1990)}]{choi1990}%
	\BibitemOpen
	\bibfield  {author} {\bibinfo {author} {\bibfnamefont {G.~M.}\ \bibnamefont
			{Choi}}\ and\ \bibinfo {author} {\bibfnamefont {H.~L.}\ \bibnamefont
			{Tuller}},\ }\bibfield  {title} {\bibinfo {title} {Nonstoichiometry and mixed
			conduction in $\alpha$‐{Ta$_2$O$_5$}},\ }\href
	{https://doi.org/10.1111/j.1151-2916.1990.tb09815.x} {\bibfield  {journal}
		{\bibinfo  {journal} {J. Am. Ceram. Soc.}\ }\textbf {\bibinfo {volume}
			{73}},\ \bibinfo {pages} {1700} (\bibinfo {year} {1990})}\BibitemShut
	{NoStop}%
	\bibitem [{\citenamefont {Xu}\ \emph {et~al.}(2012)\citenamefont {Xu},
		\citenamefont {Ding},\ and\ \citenamefont {Ma}}]{xu2012}%
	\BibitemOpen
	\bibfield  {author} {\bibinfo {author} {\bibfnamefont {M.}~\bibnamefont
			{Xu}}, \bibinfo {author} {\bibfnamefont {J.}~\bibnamefont {Ding}},\ and\
		\bibinfo {author} {\bibfnamefont {E.}~\bibnamefont {Ma}},\ }\bibfield
	{title} {\bibinfo {title} {One-dimensional stringlike cooperative migration
			of lithium ions in an ultrafast ionic conductor},\ }\href
	{https://doi.org/10.1063/1.4737397} {\bibfield  {journal} {\bibinfo
			{journal} {Appl. Phys. Lett.}\ }\textbf {\bibinfo {volume} {101}},\ \bibinfo
		{pages} {031901} (\bibinfo {year} {2012})}\BibitemShut {NoStop}%
	\bibitem [{\citenamefont {Deng}\ \emph {et~al.}(2015)\citenamefont {Deng},
		\citenamefont {Eames}, \citenamefont {Chotard}, \citenamefont {Lal{\`e}re},
		\citenamefont {Seznec}, \citenamefont {Emge}, \citenamefont {Pecher},
		\citenamefont {Grey}, \citenamefont {Masquelier},\ and\ \citenamefont
		{Islam}}]{deng2015}%
	\BibitemOpen
	\bibfield  {author} {\bibinfo {author} {\bibfnamefont {Y.}~\bibnamefont
			{Deng}}, \bibinfo {author} {\bibfnamefont {C.}~\bibnamefont {Eames}},
		\bibinfo {author} {\bibfnamefont {J.-N.}\ \bibnamefont {Chotard}}, \bibinfo
		{author} {\bibfnamefont {F.}~\bibnamefont {Lal{\`e}re}}, \bibinfo {author}
		{\bibfnamefont {V.}~\bibnamefont {Seznec}}, \bibinfo {author} {\bibfnamefont
			{S.}~\bibnamefont {Emge}}, \bibinfo {author} {\bibfnamefont {O.}~\bibnamefont
			{Pecher}}, \bibinfo {author} {\bibfnamefont {C.~P.}\ \bibnamefont {Grey}},
		\bibinfo {author} {\bibfnamefont {C.}~\bibnamefont {Masquelier}},\ and\
		\bibinfo {author} {\bibfnamefont {M.~S.}\ \bibnamefont {Islam}},\ }\bibfield
	{title} {\bibinfo {title} {Structural and mechanistic insights into fast
			lithium-ion conduction in li4sio4--li3po4 solid electrolytes},\ }\href
	{https://doi.org/10.1021/jacs.5b04444} {\bibfield  {journal} {\bibinfo
			{journal} {J. Am. Chem. Soc.}\ }\textbf {\bibinfo {volume} {137}},\ \bibinfo
		{pages} {9136} (\bibinfo {year} {2015})}\BibitemShut {NoStop}%
	\bibitem [{\citenamefont {He}\ \emph {et~al.}(2017)\citenamefont {He},
		\citenamefont {Zhu},\ and\ \citenamefont {Mo}}]{he2017}%
	\BibitemOpen
	\bibfield  {author} {\bibinfo {author} {\bibfnamefont {X.}~\bibnamefont
			{He}}, \bibinfo {author} {\bibfnamefont {Y.}~\bibnamefont {Zhu}},\ and\
		\bibinfo {author} {\bibfnamefont {Y.}~\bibnamefont {Mo}},\ }\bibfield
	{title} {\bibinfo {title} {Origin of fast ion diffusion in super-ionic
			conductors},\ }\href {https://doi.org/10.1038/ncomms15893} {\bibfield
		{journal} {\bibinfo  {journal} {Nat. Commun.}\ }\textbf {\bibinfo {volume}
			{8}},\ \bibinfo {pages} {15893} (\bibinfo {year} {2017})}\BibitemShut
	{NoStop}%
	\bibitem [{\citenamefont {Kresse}\ and\ \citenamefont
		{Furthm\"uller}(1996)}]{VASP1}%
	\BibitemOpen
	\bibfield  {author} {\bibinfo {author} {\bibfnamefont {G.}~\bibnamefont
			{Kresse}}\ and\ \bibinfo {author} {\bibfnamefont {J.}~\bibnamefont
			{Furthm\"uller}},\ }\bibfield  {title} {\bibinfo {title} {Efficient iterative
			schemes for ab initio total-energy calculations using a plane-wave basis
			set},\ }\href {https://doi.org/10.1103/PhysRevB.54.11169} {\bibfield
		{journal} {\bibinfo  {journal} {Phys. Rev. B}\ }\textbf {\bibinfo {volume}
			{54}},\ \bibinfo {pages} {11169} (\bibinfo {year} {1996})}\BibitemShut
	{NoStop}%
	\bibitem [{\citenamefont {Kresse}\ and\ \citenamefont {Joubert}(1999)}]{VASP2}%
	\BibitemOpen
	\bibfield  {author} {\bibinfo {author} {\bibfnamefont {G.}~\bibnamefont
			{Kresse}}\ and\ \bibinfo {author} {\bibfnamefont {D.}~\bibnamefont
			{Joubert}},\ }\bibfield  {title} {\bibinfo {title} {From ultrasoft
			pseudopotentials to the projector augmented-wave method},\ }\href
	{https://doi.org/10.1103/PhysRevB.59.1758} {\bibfield  {journal} {\bibinfo
			{journal} {Phys. Rev. B}\ }\textbf {\bibinfo {volume} {59}},\ \bibinfo
		{pages} {1758} (\bibinfo {year} {1999})}\BibitemShut {NoStop}%
	\bibitem [{\citenamefont {Bl\"ochl}(1994)}]{PAW}%
	\BibitemOpen
	\bibfield  {author} {\bibinfo {author} {\bibfnamefont {P.~E.}\ \bibnamefont
			{Bl\"ochl}},\ }\bibfield  {title} {\bibinfo {title} {Projector augmented-wave
			method},\ }\href {https://doi.org/10.1103/PhysRevB.50.17953} {\bibfield
		{journal} {\bibinfo  {journal} {Phys. Rev. B}\ }\textbf {\bibinfo {volume}
			{50}},\ \bibinfo {pages} {17953} (\bibinfo {year} {1994})}\BibitemShut
	{NoStop}%
	\bibitem [{\citenamefont {Perdew}\ \emph {et~al.}(2008)\citenamefont {Perdew},
		\citenamefont {Ruzsinszky}, \citenamefont {Csonka}, \citenamefont {Vydrov},
		\citenamefont {Scuseria}, \citenamefont {Constantin}, \citenamefont {Zhou},\
		and\ \citenamefont {Burke}}]{pbesol}%
	\BibitemOpen
	\bibfield  {author} {\bibinfo {author} {\bibfnamefont {J.~P.}\ \bibnamefont
			{Perdew}}, \bibinfo {author} {\bibfnamefont {A.}~\bibnamefont {Ruzsinszky}},
		\bibinfo {author} {\bibfnamefont {G.~I.}\ \bibnamefont {Csonka}}, \bibinfo
		{author} {\bibfnamefont {O.~A.}\ \bibnamefont {Vydrov}}, \bibinfo {author}
		{\bibfnamefont {G.~E.}\ \bibnamefont {Scuseria}}, \bibinfo {author}
		{\bibfnamefont {L.~A.}\ \bibnamefont {Constantin}}, \bibinfo {author}
		{\bibfnamefont {X.}~\bibnamefont {Zhou}},\ and\ \bibinfo {author}
		{\bibfnamefont {K.}~\bibnamefont {Burke}},\ }\bibfield  {title} {\bibinfo
		{title} {Restoring the density-gradient expansion for exchange in solids and
			surfaces},\ }\href {https://doi.org/10.1103/PhysRevLett.100.136406}
	{\bibfield  {journal} {\bibinfo  {journal} {Phys. Rev. Lett.}\ }\textbf
		{\bibinfo {volume} {100}},\ \bibinfo {pages} {136406} (\bibinfo {year}
		{2008})}\BibitemShut {NoStop}%
	\bibitem [{\citenamefont {Nosé}(1991)}]{md1}%
	\BibitemOpen
	\bibfield  {author} {\bibinfo {author} {\bibfnamefont {S.}~\bibnamefont
			{Nosé}},\ }\bibfield  {title} {\bibinfo {title} {Constant temperature
			molecular dynamics methods},\ }\href {https://doi.org/10.1143/PTPS.103.1}
	{\bibfield  {journal} {\bibinfo  {journal} {Prog. Theor. Phys. Supp.}\
		}\textbf {\bibinfo {volume} {103}},\ \bibinfo {pages} {1} (\bibinfo {year}
		{1991})}\BibitemShut {NoStop}%
	\bibitem [{\citenamefont {Bylander}\ and\ \citenamefont
		{Kleinman}(1992)}]{md2}%
	\BibitemOpen
	\bibfield  {author} {\bibinfo {author} {\bibfnamefont {D.~M.}\ \bibnamefont
			{Bylander}}\ and\ \bibinfo {author} {\bibfnamefont {L.}~\bibnamefont
			{Kleinman}},\ }\bibfield  {title} {\bibinfo {title} {Energy fluctuations
			induced by the nos\'e thermostat},\ }\href
	{https://doi.org/10.1103/PhysRevB.46.13756} {\bibfield  {journal} {\bibinfo
			{journal} {Phys. Rev. B}\ }\textbf {\bibinfo {volume} {46}},\ \bibinfo
		{pages} {13756} (\bibinfo {year} {1992})}\BibitemShut {NoStop}%
	\bibitem [{\citenamefont {Togo}\ and\ \citenamefont {Tanaka}(2015)}]{phonopy}%
	\BibitemOpen
	\bibfield  {author} {\bibinfo {author} {\bibfnamefont {A.}~\bibnamefont
			{Togo}}\ and\ \bibinfo {author} {\bibfnamefont {I.}~\bibnamefont {Tanaka}},\
	}\bibfield  {title} {\bibinfo {title} {First principles phonon calculations
			in materials science},\ }\href
	{https://doi.org/https://doi.org/10.1016/j.scriptamat.2015.07.021} {\bibfield
		{journal} {\bibinfo  {journal} {Scr. Mater.}\ }\textbf {\bibinfo {volume}
			{108}},\ \bibinfo {pages} {1} (\bibinfo {year} {2015})}\BibitemShut {NoStop}%
	\bibitem [{\citenamefont {Momma}\ and\ \citenamefont {Izumi}(2011)}]{vesta}%
	\BibitemOpen
	\bibfield  {author} {\bibinfo {author} {\bibfnamefont {K.}~\bibnamefont
			{Momma}}\ and\ \bibinfo {author} {\bibfnamefont {F.}~\bibnamefont {Izumi}},\
	}\bibfield  {title} {\bibinfo {title} {{{\it VESTA3} for three-dimensional
				visualization of crystal, volumetric and morphology data}},\ }\href
	{https://doi.org/10.1107/S0021889811038970} {\bibfield  {journal} {\bibinfo
			{journal} {J. Appl. Cryst.}\ }\textbf {\bibinfo {volume} {44}},\ \bibinfo
		{pages} {1272} (\bibinfo {year} {2011})}\BibitemShut {NoStop}%
	\bibitem [{\citenamefont {Henkelman}\ \emph {et~al.}(2000)\citenamefont
		{Henkelman}, \citenamefont {Uberuaga},\ and\ \citenamefont
		{Jónsson}}]{neb1}%
	\BibitemOpen
	\bibfield  {author} {\bibinfo {author} {\bibfnamefont {G.}~\bibnamefont
			{Henkelman}}, \bibinfo {author} {\bibfnamefont {B.~P.}\ \bibnamefont
			{Uberuaga}},\ and\ \bibinfo {author} {\bibfnamefont {H.}~\bibnamefont
			{Jónsson}},\ }\bibfield  {title} {\bibinfo {title} {{A climbing image nudged
				elastic band method for finding saddle points and minimum energy paths}},\
	}\href {https://doi.org/10.1063/1.1329672} {\bibfield  {journal} {\bibinfo
			{journal} {J. Chem. Phys.}\ }\textbf {\bibinfo {volume} {113}},\ \bibinfo
		{pages} {9901} (\bibinfo {year} {2000})}\BibitemShut {NoStop}%
	\bibitem [{\citenamefont {Henkelman}\ and\ \citenamefont
		{Jónsson}(2000)}]{neb2}%
	\BibitemOpen
	\bibfield  {author} {\bibinfo {author} {\bibfnamefont {G.}~\bibnamefont
			{Henkelman}}\ and\ \bibinfo {author} {\bibfnamefont {H.}~\bibnamefont
			{Jónsson}},\ }\bibfield  {title} {\bibinfo {title} {{Improved tangent
				estimate in the nudged elastic band method for finding minimum energy paths
				and saddle points}},\ }\href {https://doi.org/10.1063/1.1323224} {\bibfield
		{journal} {\bibinfo  {journal} {J. Chem. Phys.}\ }\textbf {\bibinfo {volume}
			{113}},\ \bibinfo {pages} {9978} (\bibinfo {year} {2000})}\BibitemShut
	{NoStop}%
\end{thebibliography}
%

\newpage

\begin{figure*}[p]
	\hspace*{-0.5cm} \vspace*{-.2cm}	\includegraphics[scale=0.92]{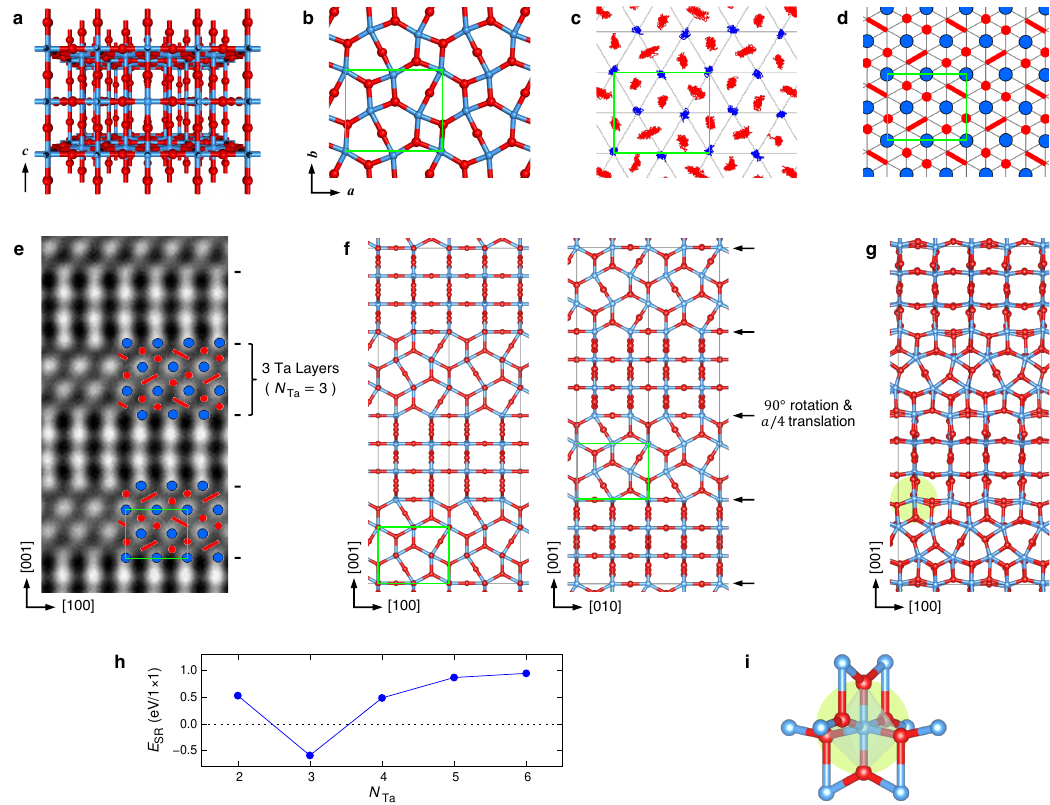}
	\newpage
	\caption{{\bf Orthorhombic-$\lambda$ structure of L-Ta$_\mathbf{2}$O$_\mathbf{5}$ vs.\ tetragonal-$\lambda$ structure of H-Ta$_\mathbf{2}$O$_\mathbf{5}$.}
		{\bf a-d}~Orthorhombic-$\lambda$ structure of L-Ta$_2$O$_5$ \cite{lee2013}.
		{\bf a}~Perspective view; blue and red spheres represent Ta and O atoms, respectively.
		{\bf b}~Coordination environment within a Ta$_2$O$_3$ layer, with the unit cell outlined in green.
		{\bf c}~Triangular arrangement of Ta atoms (projected in-plane) obtained from molecular dynamics at finite temperature, consistent with the triangular Ta sublattice observed in X-ray diffraction \cite{lehovec1964}.
		{\bf d}~Adaptation of the orthorhombic-$\lambda$ model onto a triangular grid.
		{\bf e-i}~Tetragonal-$\lambda$ structure of H-Ta$_2$O$_5$.
		{\bf e}~TEM image of the Ta sublattice (viewed along the [010]-axis) showing the tetragonal symmetry (\emph{I$4_1$/amd}), reproduced from ref.~\cite{liu2007} with permission, overlaid with the triangular-grid orthorhombic-$\lambda$ model. $N_{\rm Ta}$ denotes the number of Ta atom rows between 90° screw-rotation planes.
		{\bf f}~Initial (pre-DFT-relaxation) atomic model of H-Ta$_2$O$_5$ constructed by inserting the orthorhombic-$\lambda$ structure into a tetragonal lattice with right-handed 4$_1$ screw axes (space group \emph{P$4_1 2_1 2$}). The model is shown in two projections; black arrows mark the screw-rotation planes.
		{\bf g}~Fully relaxed atomic structure of the model in (f).
		{\bf h}~$E_{\rm SR}$, the energy cost of the orthorhombic-$\lambda$ to tetragonal-$\lambda$ transformation (see Methods), versus $N_{\rm Ta}$ (Fig.~S3). $E_{\rm SR}$ is negative solely at $N_{\rm Ta}=3$.
		{\bf i}~Ta coordination at screw-rotation plane. Four equatorial O atoms exhibit vertical flexibility, a key feature enabling collective oxygen transport.
	}
\end{figure*}

\begin{figure*}[p]
	\hspace*{-0.4cm}
	\includegraphics[scale=1.0]{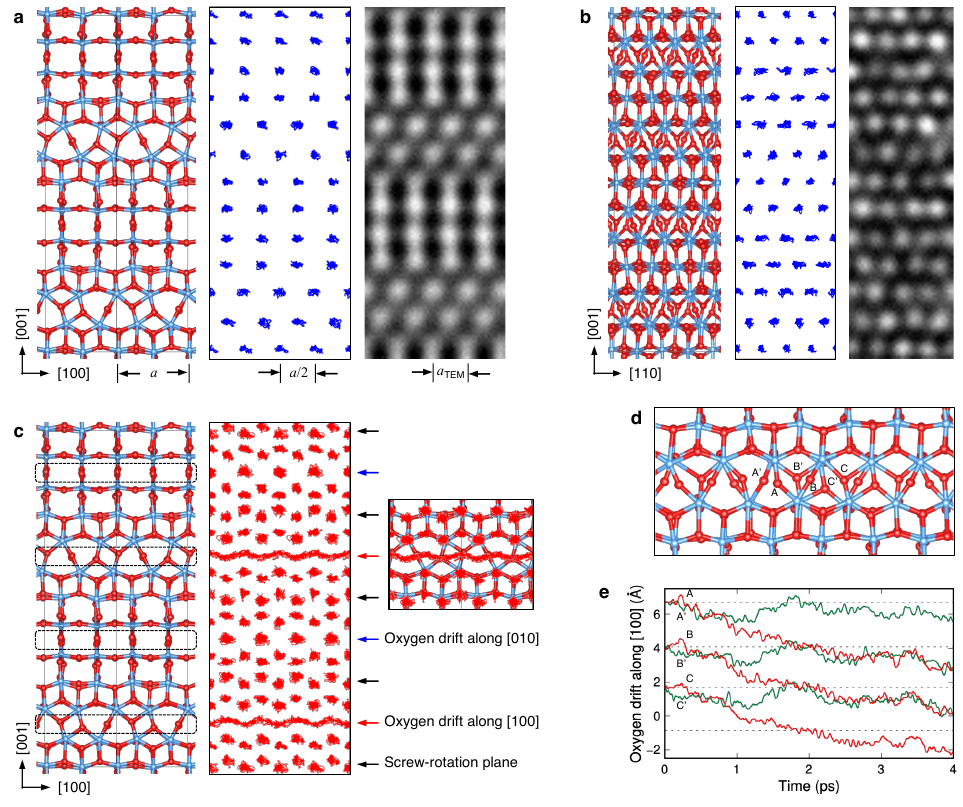}
	\caption{{\bf Dynamic behavior of the tetragonal-$\lambda$ structure at 1000 K from molecular dynamics simulations.}
		{\bf a}~Viewed along the [$010$] direction: (left) static atomic structure, (center) Ta atom trajectories at 1000 K, and (right) TEM image reproduced from ref.~\cite{liu2007} with permission. Thermal motion of Ta atoms creates a positional `cloud' around each site, forming a blurred triangular lattice in the Ta$_2$O$_3$ plane.
		{\bf b}~Viewed along the [$1\overline{1}0$] direction. The time-averaged Ta arrangement aligns well with the Ta sublattice in the corresponding TEM image (reproduced from ref.~\cite{liu2007} with permission).
		{\bf c}~O atom trajectories at 1000 K (center). Most O atoms remain close to their lattice sites (red clouds), while a subset, located midway between screw-rotation planes (right), exhibit long, wavy paths, indicating collective diffusion within the layer. This collective drift is confined to each Ta$_2$O$_3$ layer and alternates between the [100] and [010] directions along [001], as shown in the left panel (dashed-line boxes).
		{\bf d}~Snapshot from the MD simulation, showing two adjacent Ta$_2$O$_3$ layers.
		{\bf e}~Displacement of three consecutive migrating O atoms (labeled in panel d) along [100] over time, illustrating their cooperative motion. The migrating O atoms maintain their relative ordering and spacing.
	}
\end{figure*}

\begin{figure*}[t]
	\hspace*{-0.3cm}
	\includegraphics[scale=1.0]{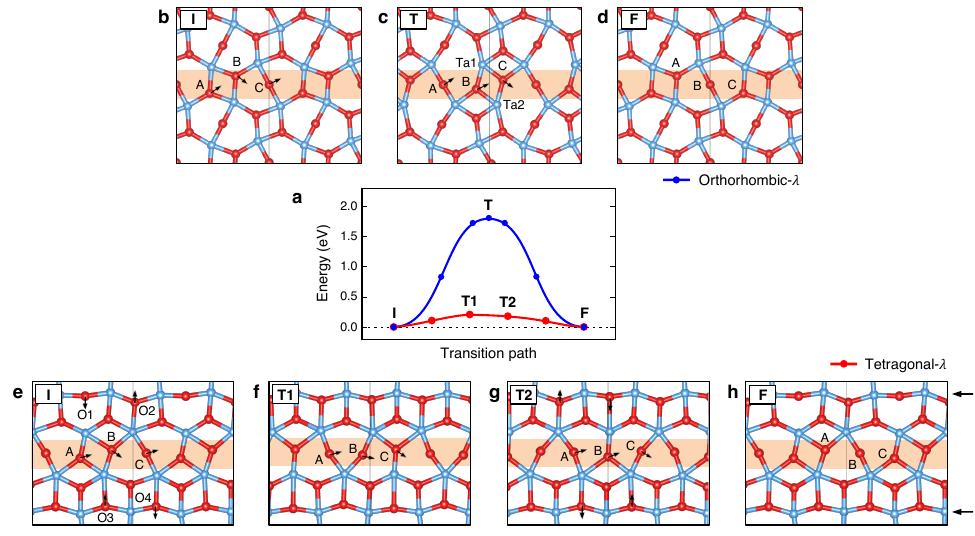}
	\caption{{\bf Energy barrier and transition state for cooperative oxygen migration.}
		{\bf a}~Nudged elastic band (NEB) energy profiles comparing the oxygen migration pathway in the orthorhombic-$\lambda$ and tetragonal-$\lambda$ structures. The tetragonal phase exhibits a significantly lower migration barrier (0.21 eV) compared to the orthorhombic phase (1.8 eV), indicating facile oxygen transport even without defects.
		{\bf b–d}~Atomic configurations of the initial (I), transition (T), and final (F) states for oxygen migration in the orthorhombic-$\lambda$ structure.
		{\bf e–h}~Atomic configurations of the initial (I), transition (T1 and T2), and final (F) states for oxygen migration in the tetragonal-$\lambda$ structure.
		In both structures, three oxygen ions (A, B, C) move cooperatively between equivalent lattice sites, resulting in identical initial and final states, demonstrating a cyclic, collective migration mechanism.
		In the orthorhombic-$\lambda$ transition state, Ta1 and Ta2 atoms exhibit significant bond distortions. Conversely, in the tetragonal-$\lambda$ transition state, flexible vertical motion of O atoms at the screw-rotation planes (O1 to O4) minimizes bond distortions at Ta sites, thereby lowering the oxygen migration energy barrier.
	}
\end{figure*}

\begin{figure*}[t]
	\includegraphics[scale=1.0]{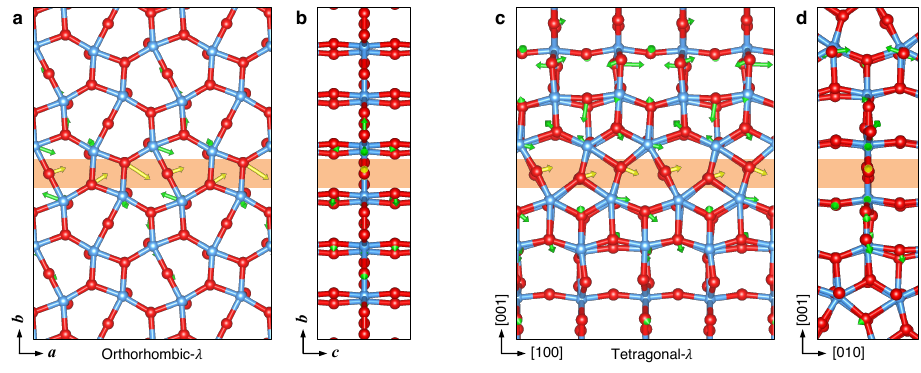}
	\caption{{\bf Structural relaxation accompanying collective oxygen motion.}
		{\bf a,b}~Displacement vectors from the initial (I) to the transition (T) state in the orthorhombic-$\lambda$ structure, and {\bf c,d}~from the initial (I) to the transition (T1) state in the tetragonal-$\lambda$ structure. Arrows originate at initial atomic positions and point to their locations in the transition state, illustrating how the lattice rearranges around the migrating oxygen. Yellow arrows (drawn to scale) trace the paths of drifting oxygen ions, while green arrows (magnified 4× for clarity) show the smaller shifts of other atoms. The tetragonal lattice exhibits extensive relaxation of the surrounding framework to accommodate the oxygen motion, in contrast to the more rigid response of the orthorhombic lattice. This enhanced structural flexibility in the tetragonal phase helps relieve local strain during oxygen migration, thereby lowering the energy barrier for transport.
	}
\end{figure*}

\begin{figure*}[t]
	\includegraphics[scale=1.0]{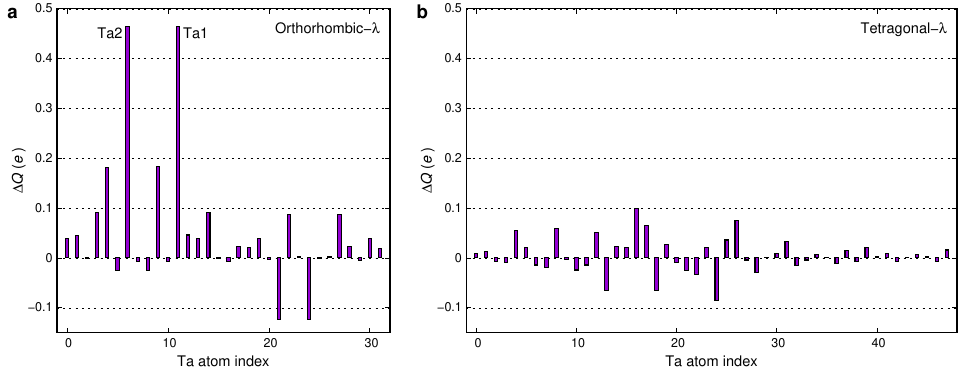}
	\caption{{\bf Dynamic charge redistribution at the oxygen migration transition state.}
		{\bf a}~Change in Ta atomic charges ($\Delta Q$) upon reaching the transition state in the orthorhombic-$\lambda$ structure.
		{\bf b}~$\Delta Q$ for each Ta atom in the tetragonal-$\lambda$ structure under the same conditions. 
		Atomic charges are computed by projecting the Kohn-Sham eigenstates onto atomic spherical harmonics within a sphere of radius 1.5 \AA, as implemented in VASP. A positive $\Delta Q$ indicates an increase in electron density on that Ta atom from the initial state to the transition state.
		The tetragonal phase shows only minor charge shifts ($|\Delta Q| < 0.1e$) distributed across many Ta atoms, indicating that electron density is flexibly redistributed to compensate for the moving oxygen. In the orthorhombic phase, larger localized charge changes occur on specific Ta atoms (Ta1 and Ta2, as labeled in Fig.~3c), reflecting less effective charge redistribution. This ability of the tetragonal lattice to dynamically spread out and neutralize charge perturbations stabilizes the transition state and complements the structural relaxation mechanism (Fig. 4) in facilitating collective ionic transport.
	}
\end{figure*}

\end{document}